\documentstyle[preprint,prb,aps,eqsecnum]{revtex}
\begin{document}
\draft
\title{Canted antiferromagnetic and spin singlet
quantum Hall states in double-layer
 systems}
\author{S. Das Sarma${}^1$, Subir Sachdev${}^2$, and Lian Zheng${}^1$}
\address{$^1$Department of Physics,
University of Maryland, College Park, Maryland 20742-4111}
\address{$^2$Department of Physics,
Yale University, P.O. Box 208120,
New Haven, CT 06520-8120
}
\date{\today}
\maketitle
\begin{abstract}
We present details of earlier studies (Zheng {\em et al},
Phys. Rev. Lett. {\bf 78}, 310 (1997) and Das Sarma {\em et al},
{\em ibid} {\bf 79}, 917 (1997)) 
and additional new results
on double-layer quantum Hall systems at a total filling
$\nu=2 \nu_1$, where a single layer at filling $\nu_1$ forms a ferromagnetic,
fully
spin-polarized, gapped incompressible
quantum Hall state. For the case $\nu_1 = 1$,
a detailed Hartree-Fock analysis is carried out on a realistic,
microscopic Hamiltonian.
Apart from the state continuously connected to the ground state of
two well separated layers,
we find two double-layer
quantum Hall phases: one
with a finite interlayer
antiferromagnetic spin ordering in the plane orthogonal to the applied
field (the `canted' state), and the other a spin singlet.
The quantum transitions between the various quantum Hall states are
continuous, and are signaled by the softening of
collective intersubband spin density excitations.
For the case of general $\nu_1$, closely related results
are obtained by a semi-phenomenological continuum
quantum field theory description of the low-lying spin
excitations using a non-linear sigma model.
Because of its broken symmetry,
the canted phase supports a
linearly dispersing Goldstone mode
and has a finite temperature Kosterlitz-Thouless transition.
We present results on the form of the phase diagram,
the magnitude of the canted order parameter,
the collective excitation dispersions,
the specific heat, the form of the dynamic light scattering spectrum at
finite temperature,
and the Kosterlitz-Thouless critical temperature.
Our findings are consistent with recent experimental results.
\end{abstract}
\pacs{73.40.Hm 73.20.Mf 73.20.Dx }
\narrowtext
\section{Introduction}
\label{intr}
Interaction in a low-dimensional
system
does not merely result in strong
renormalization of physical quantities, but can in many cases
drive the system into completely new
phases with peculiar properties.
For a two-dimensional (2D)
electron gas in a perpendicular
magnetic field, the interaction effects are especially important
because of Landau level quantization.
When electrons are entirely restricted
to the lowest Landau level by a large magnetic field,
electron-electron interaction completely dominates
the properties of the system as the electron
kinetic energy is
quenched to an unimportant constant.
One of the most interesting phenomena in this strongly-correlated system
is the quantum Hall (QH) effect, which has attracted
a great deal of experimental and theoretical interest
during the last fifteen years\cite{smg1}.
Recent advances in materials growth techniques have
made it possible to fabricate high-quality double-layer
two-dimensional electron systems with the electrons confined
to two parallel planes separated by a distance comparable to that
between electrons within a plane.
With the introduction of this
layer degree of freedom, many qualitatively new
effects due entirely to interlayer correlations
appear\cite{kun1,girvin1,girvin2,nuh,pit1,we1,ap1,dsz1}.
Many new QH phases in double-layer systems become real possibilities
because of the increased degree of freedom and the
complicated interplay among
interlayer tunneling energy, Zeeman energy, and
electron-electron Coulomb interaction energy.

In this paper, we present the details of our earlier theoretical
investigations\cite{we1,dsz1} of
the possible QH phases in a double-layer system
at a {\it total} Landau level filling factor
$\nu=2\nu_1$, where $\nu_1$ is a filling factor at which an isolated
single layer system forms a fully spin polarized incompressible
QH state ({\it e.g. $\nu_1=1,\ 1/3$, etc.})
We will discuss three distinct ground states, and the nature of
the zero or finite temperature transitions/crossovers between them:
\begin{itemize}
\item
A fully polarized ferromagnetic (FPF) 
QH state in which the spins in each layer are
aligned parallel to the magnetic field. This state is
adiabatically connected to the ground state of well separated
layers, each forming a polarized QH state at filling fraction $\nu_1$.
We will denote this FPF state also 
as the FM (for ``ferromagnetic'') state.
\item
A spin singlet (SS) state, which can be visualized crudely
as consisting of singlet pairs of electrons in opposite
layers.
Alternatively, at $\nu_1 = 1$, we will discuss the Hartree-Fock
picture of spin up and spin down electrons fully occupying
single-particle states which are symmetric in the layer
``pseudospin'' index; hence the 
singlet state will also be referred to as
SYM. In the limit of a vanishing tunneling matrix element between
the layers, this state is simply the pseudospin polarized state
of Refs~\onlinecite{girvin1,girvin2} for both spin up and
spin down electrons separately.
Throughout, we will consider the case of a non-vanishing tunneling
matrix element: in this case the pseudospin polarization is chosen
by the phase of the tunneling amplitude, and not spontaneously.
None of the phase transitions we consider here require a vanishing
tunneling matrix element; on the contrary, changes in the value
of the tunneling matrix element can drive the quantum transitions.
\item
A canted state (C) in which the average spin moments in the layers have an
antiferromagnetic
correlation in the plane perpendicular to the magnetic field,
and a ferromagnetic correlation parallel to the magnetic field.
Both ferromagnetic and antiferromagnetic moments can vary
continuously at zero temperature as parameters are varied.
The planar antiferromagnetic ordering breaks spin rotation
symmetry about the magnetic field axis:
as a consequence there is a gapless, linearly dispersing, Goldstone collective
mode in its excitation spectrum
and a Kosterlitz-Thouless transition at a finite temperature.
The C phase is the canted antiferromagnetic phase (CAF)
discussed in our earlier short publications \cite{we1,dsz1}.
\end{itemize}

We will use two distinct and complementary approaches to
understand these phases. The first is a mean-field
Hartree-Fock calculation: this applies only for integer
values of $\nu_1$, but has the advantage of working with a
precise microscopic Hamiltonian involving only parameters
which are directly known experimentally. The second
is a phenomenological, quantum field-theoretic formulation
which applies for general $\nu_1$, and allows us to more precisely
understand the consequences of thermal and quantum fluctuations.
We will now discuss some of the results of these two approaches
in turn.

In the Hartree-Fock approximation\cite{wig1},
we are able to show that the canted antiferromagnetic (C) phase is the
energetically favored ground state for $\nu=2$ at
intermediate layer separations for systems
with small Zeeman energy, and that the phase transitions from the
C to the FM or SYM phases are continuous.
We evaluate at $\nu=2$ the intersubband spin density wave (SDW)
dispersions of all phases in
the time-dependent Hartree-Fock approximation
\cite{kal1} and show that, as the precursor of the phase transitions,
the collective intersubband SDW mode softens at the phase boundaries of the
FM and SYM phases to the C phase.
The SDW becomes the linearly dispersing Goldstone mode
in the C phase, and the temperature of the Kosterlitz-Thouless
transition is obtained by evaluating its
effective spin-stiffness in the Hartree-Fock approximation.
In addition we present results on the stability energetics of the various
phases, the
antiferromagnetic order parameter, the phase diagram,
the collective intersubband SDW excitation dispersions,
and the specific heat.

The $\nu=2$ Hartree-Fock results
may also be qualitatively applicable to the
case of $\nu=6$ if the Landau level mixing is ignored
(the Landau level mixing may not be negligible at $\nu=6$, though.)
On the other hand, the situation at $\nu=4$ is very different
from the situation at $\nu=2$,
since the inter-Landau level excitation energies are
comparable to the cyclotron energy; our results do not apply at
$\nu=4$.

The microscopic Hartree-Fock analysis obviously does not apply to a
situation where the average filling factor $\nu_1$ in each layer is
fractional (e.g. $\nu_1=1/3$) with each isolated layer supporting a spin
polarized Laughlin fractional QHE state; such a 
many-body state will not
appear in any mean-field decoupling of the Hamiltonian.
However, an essential property of the phases we are discussing is
that they all have a gap towards charged excitations,
and the transitions between them are driven by changes in the
nature of the mean spin polarizations, and of the spin excitations.
This suggests that it may be possible to develop a more general
effective theory which focuses on the spin excitations alone.
We will present such a theory in Section~\ref{nls}: it turns out
to be the $O(3)$ quantum non-linear sigma model in the
presence of a magnetic field. For the case $\nu=2$, we are
able to use our earlier Hartree-Fock computations to precisely
obtain all the renormalized parameters which universally
determine the low temperature properties of the non-linear
sigma model; for other values of $\nu_1$, including the fractional
cases, these parameters remain as phenomenological inputs.
We will present the phase diagram of the sigma model, and describe
the nature of the finite temperature crossovers above the various
phases in some details. In particular, we will obtain explicit
predictions for the temperature dependence of the line
shape of the inelastic light scattering spectrum.

We note that
our findings from the two approaches are consistent with recent
inelastic light scattering measurement\cite{ap1},
where a remarkable (and temperature dependent)
softening of the long wavelength intersubband SDW mode
in a $\nu=2$ double-layer system is observed. We hope that our
other explicit theoretical results may be tested in
future experiments. The experimental situation will
be discussed in Section~\ref{exp}.

This paper is organized as follows.
The results of the Hartree-Fock theory are presented in
Section~\ref{hft}.
In Section~\ref{grou}, we study the ground state properties of the
$\nu=2$ double-layer system in a self-consistent
mean-field approximation. The intersubband SDW excitations
in the time-dependent Hartree-Fock approximation
and associated mode softening are studied
in Section~\ref{sdw}.
The thermodynamic properties are discussed in
Section~\ref{KT}, and some further discussion, along with an assessment of the
validity of the calculation, appear in Sections~\ref{multi} and~\ref{earlier}.
In a long and self-contained Section \ref{nls} we give our non-linear $\sigma$
model effective field theoretic description for a generic
$\nu=2\nu_1$ situation.
Comparison of our theory with
recent light scattering experiments is discussed
in section \ref{exp}. A short summary in section \ref{sum} concludes this paper.
We note that the readers who are interested only in microscopic Hartree-Fock
theory could skip Section \ref{nls}, and the readers who are
interested only in our long wavelength effective field theory could skip
Section~\ref{hft}. We have taken care in writing the two parts
of our work, namely the microscopic Hartree-Fock calculation for $\nu=2$
(Section \ref{hft}) and the non-linear $\sigma$ model
description for $\nu=2\nu_1$ (Section \ref{nls}) as two separate
self-contained pieces which can be read reasonably independent of
each other if so desired.

\section{Hartree-Fock Theory}
\label{hft}

We begin by writing down the explicit microscopic Hamiltonian
of a double layer quantum Hall system.

Within
the lowest Landau level,
the single particle eigenstates
may be denoted by $|\alpha\mu\sigma
\rangle$, where $\alpha$ is the intra-Landau-level index in
the lowest Landau level,
$\mu=0,1$ is the pseudospin index which
labels the symmetric and antisymmetric subbands,
and the spin index
$\sigma=\pm1$ labels $\uparrow$ and $\downarrow$ spins.\cite{rk3}
The Hamiltonian of the double-layer system is
\begin{equation}
H=H_0+H_{\rm I},
\label{equ:h0}
\end{equation}
where the non-interacting Hamiltonian is
\begin{equation}
H_0=-\Delta_{\rm sas}\sum_{\alpha\mu\sigma}(1/2-\mu) C^\dagger_{\alpha\mu\sigma}
C_{\alpha\mu\sigma}-\Delta_z\sum_{\alpha\mu\sigma}{\sigma\over2}
C^\dagger_{\alpha\mu\sigma}C_{\alpha\mu\sigma},
\label{equ:h1}
\end{equation}
where the pseudospin splitting $\Delta_{\rm sas}$ is the tunneling-induced
symmetric-antisymmetric energy separation,
the spin splitting $\Delta_z$ is the
Zeeman energy, and $C^\dagger\ (C)$ is
electron creation (annihilation) operator.
The Coulomb interaction Hamiltonian
$H_{\rm I}$ is
\begin{eqnarray}
H_{\rm I}=&&{1\over2}\sum_{\sigma_1\sigma_2}\sum_{\mu_1\mu_2\mu_3\mu_4}
\sum_{\alpha_1\alpha_2}{1\over\Omega}\sum_{\bf q}
V_{\mu_1\mu_2\mu_3\mu_4}(q)e^{-q^2l_o^2/2}e^{iq_x(\alpha_1-\alpha_2)l_o^2}
\nonumber \\
&&\times C^\dagger_{\alpha_1+q_y\mu_1\sigma_1}C^\dagger_{\alpha_2\mu_2\sigma_2}
C_{\alpha_2+q_y\mu_3\sigma_2}C_{\alpha_1\mu_4\sigma_1},
\label{equ:hi}
\end{eqnarray}
where $\Omega$ is the area of the system, $l_o=(\hbar c/eB)^{1/2}$
is the magnetic length. The non-zero Coulomb potential matrix elements
are $V_{0000}=V_{0110}=V_{1001}=V_{1111}=V_+$ and
$V_{1010}=V_{0101}=V_{1100}=V_{0011}=V_-$,
with $V_{\pm}(q)={1\over2}[v_a(q)\pm v_b(q)]$,
where
$v_a(q)={2\pi e^2\over\epsilon q}$
and $v_b(q)=v_a(q)e^{-qd}$
are the intralayer and interlayer Coulomb interaction potentials,
respectively. (The finite well-thickness
corrections can be taken into consideration
by including appropriate form factors\cite{wig1}.)

The following subsections will examine various properties of
$H$ at $\nu=2$ by mean-field and RPA-like treatments of the interactions
in $H_{\rm I}$.

\subsection{Ground states}
\label{grou}
In this subsection, we investigate the ground state properties of
$H$, and obtain the three phases discussed in the Introduction.
Performing Hartree-Fock pairing of (\ref{equ:hi}),
one obtains the mean-field interaction
Hamiltonian as
\begin{equation}
H_{\rm I}^{\rm HF}
=-\sum_{\sigma_1\sigma_2}\sum_{\mu_1\mu_2}X_{\mu_1\mu_2\sigma_1
\sigma_2}C^\dagger_{\mu_1\sigma_1}C_{\mu_2\sigma_2},
\label{equ:hhf1}
\end{equation}
where $X_{\mu_1\mu_2\sigma_1\sigma_2}={1\over2\pi l_o^2}
\sum_{\mu_3\mu_4}\sum_{\bf q}
V_{\mu_3\mu_1\mu_4\mu_2}(q)e^{-q^2l_o^2/2}<C^\dagger_{\mu_3\sigma_2}
C_{\mu_4\sigma_1}>$, which depends on the electronic state
being sought through the expectation value $<C^\dagger_{\mu_3\sigma_2}
C_{\mu_4\sigma_1}>$. We self-consistently search for the
symmetry broken states where,
in addition to $<C^\dagger_{\mu\sigma}C_{\mu\sigma}>\ne0$,
the possibility that $<C^\dagger_{\mu\uparrow}C_{1-\mu\downarrow}>\ne0$
is also allowed.
Because of the complete Landau level degeneracy,
the Hartree-Fock Hamiltonian $H^{\rm HF}=H_0+H_{\rm I}^{\rm HF}$
in a uniform state is a $4\times4$
matrix, representing the dimension of the subspace associated
with the spin and layer degrees of freedom.
It thus has four eigenenergies $\varepsilon_{i\pm}$ and four eigenstates
$\phi_{i\pm}$ ($i=1,2$), which are obtained as shown below.
In the non-interacting base
$(|0\uparrow\rangle,|1\downarrow\rangle,|0\downarrow
\rangle,|1\uparrow\rangle)$,
$H^{\rm HF}$ becomes

\begin{equation}
H^{\rm HF}=
\left(\begin{array}{clcr}
E_1&\Delta_1&0&0\\
\Delta_1&E_2&0&0\\
0&0&E_3&\Delta_2\\
0&0&\Delta_2&E_4
\end{array}\right),
\label{equ:hm1}
\end{equation}

where

\begin{eqnarray}
E_1=&&-{\Delta_{\rm sas}+\Delta_z\over2}-U_+(n_{1+}\sin^2{\theta_1\over2}+
n_{1-}\cos^2{\theta_1\over2})
\nonumber \\ &&
-U_-(n_{2+}\cos^2{\theta_2\over2}+
n_{2-}\sin^2{\theta_2\over2}),
\nonumber \\
E_2=&&{\Delta_{\rm sas}+\Delta_z\over2}-U_+(n_{1+}\cos^2{\theta_1\over2}+
n_{1-}\sin^2{\theta_1\over2})
\nonumber \\ &&
-U_-(n_{2+}\sin^2{\theta_2\over2}+
n_{2-}\cos^2{\theta_2\over2}),
\nonumber \\
E_3=&&{\Delta_z-\Delta_{\rm sas}\over2}-U_+(n_{2+}\sin^2{\theta_2\over2}
+n_{2-}\cos^2{\theta_2\over2})
\nonumber \\ &&
-U_-(n_{1+}\cos^2{\theta_1\over2}+
n_{1-}\sin^2{\theta_1\over2}),
\nonumber \\
E_4=&&{\Delta_{\rm sas}-\Delta_z\over2}-U_+(n_{2+}\cos^2{\theta_2\over2}+
n_{2-}\sin^2{\theta_2\over2})
\nonumber \\ &&
-U_-(n_{1+}\sin^2{\theta_1\over2}+
n_{1-}\cos^2{\theta_1\over2}),
\nonumber \\
\Delta_1=&&U_+{n_{1-}-n_{1+}\over2}\sin\theta_1+U_-
{n_{2-}-n_{2+}\over2}\sin\theta_2,
\nonumber \\
\Delta_2=&&U_+{n_{2-}-n_{2+}\over2}\sin\theta_2+U_-
{n_{1-}-n_{1+}\over2}\sin\theta_1.
\label{equ:e01}
\end{eqnarray}
where $\theta_1$ and $\theta_2$
are associated with the Hartree-Fock eigenstates $\phi_{i\pm}$
which need to be obtained self-consistently,
$n_{i\pm}$ are electron occupation numbers
$\langle\phi^\dagger_{i\pm}\phi_{i\pm}\rangle$,
and $U_{\pm}
={1\over\Omega}\sum_{\bf p}e^{-p^2l_o^2/2}V_{\pm}(p)$.
The off-diagonal matrix elements $\Delta_i$ represent the
possibility of the broken symmetry ($\langle C^\dagger_{\mu\uparrow}
C_{1-\mu\downarrow}\rangle\ne0$) mentioned above.
By diagonalizing the Hartree-Fock Hamiltonian $H^{\rm HF}$
of Eq. (\ref{equ:hm1}),
one obtains the eigenstates
\begin{equation}
(\phi_{1+},\phi_{1-},\phi_{2+},\phi_{2-})=
\left(\begin{array}{clcr}
\sin(\theta_1/2)&\cos(\theta_1/2)&0&0\ \ \ \ \ \ \\
\cos(\theta_1/2)&-\sin(\theta_1/2)&0&0\ \ \ \ \ \ \\
0&\ \ \ \ \ 0&\sin(\theta_2/2)&\cos(\theta_2/2)\\
0&\ \ \ \ \ 0&\cos(\theta_2/2)&-\sin(\theta_2/2)
\end{array}\right),
\label{equ:eig1}
\end{equation}
and the eigenenergies
\begin{eqnarray}
\varepsilon_{1\pm}=&&{E_1+E_2\over2}\pm\sqrt{{(E_1-E_2)^2\over4}+\Delta_1^2},
\nonumber \\
\varepsilon_{2\pm}=&&{E_3+E_4\over2}\pm\sqrt{{(E_3-E_4)^2\over4}+\Delta_2^2}.
\label{equ:e10}
\end{eqnarray}

Eqns. (\ref{equ:hm1}) to (\ref{equ:e10}) form the complete
self-consistent Hartree-Fock equations which need to be solved
numerically.
In fact, the only quantities to be determined
in this self-consistent manner are the two parameters
$\theta_1$ and $\theta_2$,
which, in turn, uniquely define the eigenstates through Eq. (\ref{equ:eig1}).
The eigenenergies always satisfy $\varepsilon_{i-}<\varepsilon_{j+}$
($i,j=1,2$), so the ground state at $\nu=2$ is given by
$|\rangle=\Pi_{i}\phi^\dagger_{i-}|{\rm v}\rangle$, where
$|{\rm v}\rangle$ is the vacuum state. The
ground state energy is given by $E=\langle H_0+{1\over2}H^{\rm HF}_{\rm I}
\rangle$.

There are several sets of $\theta_1$ and $\theta_2$
which make Eq. (\ref{equ:eig1}) the
self-consistent solutions to the mean-field
Hartree-Fock equations. One is $\theta_1=0$ and $\theta_2=0$,
which corresponds to the symmetric (SYM) state. Another is
$\theta_1=0$ and $\theta_2=\pi$, which corresponds to the spin polarized
ferromagnetic (FM) state.
These two are the spin-ferromagnets (FM)
or layer pseudospin-`ferromagnets' \cite{girvin1,girvin2} (SYM)
whose existence is naturally expected
in the presence of finite Zeeman and tunneling energies.
More interesting is that, for
$\Delta_{\rm sas}>\Delta_z$, there exists a
solution at intermediate interlayer separations
with $0<\theta_i<\pi$.
As we shall see shortly, this new state
possesses a canted antiferromagnetic ordering (the C phase), {\it i.e.}
an interlayer inplane antiferromagnetic spin ordering
with the inplane spin magnetic moment in each layer being
equal in magnitude and opposite of each other.
The energies of these different states are shown in Fig.\ref{fig1}.
It is clear from this figure that the energetically favored ground state
is the SYM state at small
interlayer separations, the C state
at intermediate separations, and the FM state
at large interlayer separations.
The $\nu=2$ double-layer QH system thus undergoes two
quantum phase transitions
as the layer separation is increased from $d=0$ to $d\rightarrow\infty$
at a fixed magnetic field.

To show the antiferromagnetic spin correlations,
we rearrange
the eigenstates as
\begin{equation}
\phi_{i\pm}=
(1/\sqrt{2}\ )\left(|{\rm L}
\rangle S^{\rm L}_{i\pm}+|{\rm R}\rangle S^{\rm R}_{i\pm}\right),
\label{equ:eig2}
\end{equation}
where $S^{L(R)}_{i\pm}$,
electron spin configurations in the left (right) layer in the
eigenstate $\phi_{i\pm}$, are
$S^{\rm L}_{i-}=\cos(\theta_i/2)
|\uparrow\rangle-\sin(\theta_i/2) |\downarrow\rangle$,
$S^{\rm R}_{i-}
=\cos(\theta_i/2) |\uparrow\rangle+\sin(\theta_i/2) |\downarrow\rangle$,
and satisfy $(S^{\rm L}_{i+})^\dagger S^{\rm L}_{i-}=
(S^{\rm R}_{i+})^\dagger S^{\rm R}_{i-}=0$.
We immediately obtain the canted antiferromagnetic spin order
as
\begin{equation}
\langle {\cal S}^{\rm R}_x\rangle=-\langle {\cal S}^{\rm L}
_x\rangle={1\over4}(\sin\theta_1
+\sin\theta_2),
\label{equ:op1}
\end{equation}
where ${\cal S}^{L(R)}$ is the electron
spin operator in the left (right) layer,
and $x$ denotes the spin alignment
direction within the two dimensional plane.
This canted interlayer
antiferromagnetic spin ordering is shown schematically
in Fig. \ref{fig41}.
Note that the total spin magnetic moment still points in the direction of
the magnetic field as required by symmetry.
It is obvious that this antiferromagnetic order
breaks the $U(1)$ symmetry associated with the
spin-rotational invariance of the system. Its
consequences on the low temperature thermodynamic properties will
be discussed later.
The numerical result of this order parameter
$|\langle {\cal S}^{\rm L}_x\rangle-\langle {\cal S}^{\rm R}_x\rangle|$
is shown in Fig.\ref{fig2}.
One can see that when Zeeman energy $\Delta_z$ is increased,
the range of the layer separations
where the canted antiferromagnetic state exists shrinks
in favor of the ferromagnetic state,
as the Zeeman energy obviously favors the spin polarized state.
It is clear that
the phase transition is continuous.

The phase diagram, shown in Fig. \ref{fig3},
can be constructed from this mean-field approximation.
The states $|0\uparrow\rangle$ and $|1\uparrow\rangle$ are occupied
in the FM phase, $|0\uparrow\rangle$ and $|0\downarrow\rangle$
are occupied in the SYM phase,
and the C phase interpolates between them.
The SYM phase exists for $\Delta_{\rm sas}>\Delta_z$
and $d<d_{c1}$, the C phase
exists for $\Delta_{\rm sas}>\Delta_z$
and $d_{c1}<d<d_{c2}$,
and the FM phase exists for either $\Delta_z>\Delta_{\rm sas}$
or $d>d_{c2}$.
The FM phase is favored when
$\Delta_z$ is increased, while
the SYM phase is favored when
$\Delta_{\rm sas}$ is increased.
In the next subsection, the same phase diagram
will be obtained by studying the softening of the
intersubband SDW excitations
in the time dependent Hartree-Fock approximation.

In this subsection we have studied the ground state properties
of $\nu=2$ double-layer QH systems in a mean-field Hartree-Fock
approximation and showed the existence of three stable QH phases.
The most interesting observation is the
existence of a canted antiferromagnetic phase, with a broken spin
rotation symmetry, in between the symmetric
and the ferromagnetic phases.

\subsection{Intersubband SDW excitations
and mode softening}
\label{sdw}
In this section, we study collective intersubband SDW spectrum
of $\nu=2$ double-layer QH systems in the time-dependent
Hartree-Fock approximation.\cite{kal1}
These excitations involve flipping both spin and pseudospin
of the electron and are the lowest energy excitations
at $\nu=2$.
The phase instability is studied
by investigating the softening of the collective
intersubband SDW excitations.
The results obtained in this section are in complete quantitative
agreement with
the results obtained from the ground state studies
in the previous section, as, of course, they should be if
the calculations are done correctly.

In the absence of interaction,
the two branches of the intersubband
SDW excitations which correspond to transitions
$|0\uparrow\rangle\leftrightarrow|1\downarrow\rangle$ and
$|0\downarrow\rangle\leftrightarrow|1\uparrow\rangle$
have excitation energies
$|\Delta_{\rm sas}\pm\Delta_z|$,
where $\Delta_{\rm sas}$ and $\Delta_z$
are interlayer tunneling and Zeeman energies, respectively.
Interaction renormalizes the excitation energies
in two ways. One is
a self-energy correction to the polarizability
due to the loss of exchange energy when
an electron is excited to a higher but empty level, which raises
the excitation energies. The other is
the vertex correction to the polarizability due to an excitonic attraction
between the electron excited to the higher level
and the hole it leaves behind, which lowers the excitation
energies.
In diagrammatic perturbation theories,
the effect of the exchange energy
on the excitation energies is accounted for by
including the
corresponding self-energy in electron Greens functions,
and the effect of the
excitonic attraction is represented by vertex corrections.
The self-energy and the vertex correction must be consistent with
each other obeying the Ward identity.
The direct Hartree term does not influence the SDW excitations
because Coulomb interaction is spin-rotationally invariant.
Since the Coulomb interaction potentials are subband-index dependent,
they may introduce mode-coupling between the two branches of the intersubband
SDW excitations. This mode-coupling pushes down the frequency
of the low-lying excitation
and hence helps mode softening.

The intersubband SDW excitation spectra
are obtained as the poles of the retarded intersubband spin-density response
function
\begin{equation}
\chi^{\rm ret}(q,\omega)=-i\int_0^\infty e^{i\omega t}
\langle[\rho_{SD}({\bf q},t),\rho_{SD}^\dagger(-{\bf q},0)]
\rangle,
\label{equ:x1}
\end{equation}
where the intersubband SDW operator is defined as
\begin{equation}
\rho_{SD}({\bf r})=\sum_{i=1}^2\phi^\dagger_{i-}({\bf r})\phi_{i+}({\bf r}).
\label{equ:ro1}
\end{equation}
$\rho_{SD}({\bf r})$ recovers to familiar forms
$\rho_{SD}({\bf r})=\sum_{\mu}C_{\mu\uparrow}^\dagger({\bf r})
C_{1-\mu\downarrow}({\bf r})$ in the spin polarized state
($\theta_1=0$ and $\theta_2=\pi$), and
$\rho_{SD}({\bf r})=\sum_{\sigma}C_{0\sigma}^\dagger({\bf r})
C_{1-\sigma}({\bf r})$ in the symmetric state
($\theta_1=\theta_2=0$).

$\chi^{\rm ret}(q,\omega)$
is evaluated
in the time-dependent Hartree-Fock approximation,
\cite{kal1}
which we adapt to double-layer systems
and, for simplicity, we ignore all the higher
Landau levels. As argued earlier, this should be a
good approximation for our problem.
In this approximation, one includes the single-loop self-energy and the
ladder vertex diagrams in the theory, which
satisfies the Ward identities.
This time-dependent Hartree-Fock approximation, therefore,
corresponds to solving the vertex equation shown in Fig.\ref{fig4},
where the electron propagators are the self-consistent
Hartree-Fock Green's functions obtained from the mean-field
approximation discussed in the previous section.
Due to the fact that the Coulomb interaction is frequency independent
and that the Landau levels are completely degenerate,
the integral vertex equation can be transformed into an algebraic
matrix equation.\cite{kal1}
The matrices can be further block diagonalized into $4\times4$
matrices,
from which the poles of the spin-density response function
can be (almost) analytically calculated.

Combining Eqns. (\ref{equ:x1}) and (\ref{equ:ro1}),
one obtains the spin density response function in the
Matsubara frequencies \cite{man1}
\begin{equation}
\chi(q,i\omega)=e^{-q^2l_o^2/2}\sum_{i\alpha}e^{-iq_x\alpha l_o^2}
D_{i+}(i\omega)\Gamma_{i+}(q,i\omega,\alpha),
\label{equ:x01}
\end{equation}
where
\begin{eqnarray}
D_{i\lambda}=&&{1\over\beta}\sum_n {\cal G}_{i\lambda}
(ip_n+i\omega){\cal G}_{i-\lambda}(ip_n) \nonumber \\
=&&{n_{i-\lambda}-n_{i\lambda}\over i\omega+
\varepsilon_{i-\lambda}-\varepsilon_{i\lambda}} \nonumber \\
=&&{1\over\lambda i\omega+\varepsilon_{i-}-\varepsilon_{i+}}
\ \ \  {\rm for \ \ } T=0,
\label{equ:d01}
\end{eqnarray}
where $\beta=1/k_BT$, ${\cal G}_{i\lambda}$ is the Green's function
corresponding to the self-consistent
Hartree-Fock eigenstate
$\phi_{i\lambda}$ and eigenenergy $\varepsilon_{i\lambda}$
given in Eqns. (\ref{equ:eig1}) and (\ref{equ:e10}), respectively.
The ladder diagram vertex function is
\begin{eqnarray}
\label{equ:g01}
\Gamma_{i\lambda}(q,i\omega,\alpha)&&=e^{iq_x\alpha l_o^2}
-{1\over\Omega}\sum_{p_x i^\prime\alpha^\prime\lambda^\prime}
e^{-[p_x^2+(\alpha-\alpha^\prime)^2]l_o^2/2}e^{ip_xq_yl_o^2} \\
&&\times D_{i^\prime\lambda^\prime}\Gamma_{i^\prime\lambda^\prime}
(q,i\omega,\alpha^\prime)
\langle i\lambda; i^\prime-\lambda^\prime|V(p_x,\alpha-\alpha^\prime)|
i-\lambda; i^\prime\lambda^\prime\rangle\nonumber ,
\end{eqnarray}
where the interaction matrix element is
\begin{eqnarray}
\langle i_1\lambda_1;i_2\lambda_2|V(q)|i_3\lambda_3;i_4\lambda_4\rangle
&&=
{1\over2}[1+(-1)^{i_1+i_2+i_3+i_4}]
\left((S^L_{i_1\lambda_1})^\dagger S^L_{i_4\lambda_4}\right)
\left((S^L_{i_2\lambda_2})^\dagger S^L_{i_3\lambda_3}\right) \nonumber \\
&&\times\left[V_+(q)\delta_{i_2i_3}+V_-(q)(1-\delta_{i_2i_3})\right],
\label{equ:v01}
\end{eqnarray}
where $S^L_{i\lambda}$ is the electron spin states
given in Eq. (\ref{equ:eig2}).

To solve the vertex equation, we perform
the following Fourier transformations \cite{kal1}
\begin{equation}
\overline{\Gamma}_{i\lambda}(k)=\sum_\alpha
\Gamma_{i\lambda}(\alpha)e^{-ik\alpha l_o^2},
\label{equ:g02}
\end{equation}
and
\begin{equation}
\tilde{V}_{i\lambda;i^\prime\lambda^\prime}
({\bf q})={1\over\Omega}\sum_{\bf p}e^{-p^2l_o^2/2}
e^{i{\bf p}\wedge{\bf q}l_o^2}V_{i\lambda;i^\prime\lambda^\prime}({\bf p}),
\label{equ:v02}
\end{equation}
where ${\bf p}\wedge{\bf q}=p_xq_y-p_yq_x$
and ${V}_{i\lambda;i^\prime\lambda^\prime}=
\langle i\lambda;i^\prime-\lambda^\prime|{V}(q)|
i-\lambda;i^\prime\lambda^\prime\rangle$,
as given by Eq. (\ref{equ:v01}).
After an analytical continuation,
one obtains
\begin{equation}
\chi^{\rm ret}(q,\omega)=e^{-q^2l_o^2/2}\sum_{i=1,2}
\Upsilon_{i+}(q,\omega),
\label{equ:x02}
\end{equation}
where
\begin{equation}
\Upsilon=(D^{-1}+\tilde{V})^{-1}N,
\label{equ:g03}
\end{equation}
$N$ and $\Upsilon$ are $4\times1$ matrices, with
$N_{i\lambda}=\Omega/2\pi l_o^2$, the number of magnetic flux passing
through the system,
and $\Upsilon_{i\lambda}=
D_{i\lambda}(\omega)\overline{\Gamma}_{i\lambda}(q,\omega)$.
$D$ and $\tilde{V}$ are $4\times4$ matrices,
with
$D_{i\lambda;i^\prime\lambda^\prime}=
\delta_{ii^\prime}\delta_{\lambda\lambda^\prime}
D_{i\lambda}(\omega)$,
and $\tilde{V}_{i\lambda;i^\prime\lambda^\prime}$ defined in
Eq.. (\ref{equ:v02}).

The intersubband SDW dispersion $\omega(q)$, which occurs
as the pole of the retarded spin density response function $\chi^{\rm ret}$,
is the solution to ${\rm det}|D^{-1}(\omega)+\tilde{V}(q,\omega)|=0$.
After a lengthy but straightforward algebraic manipulation,
the two intersubband SDW dispersions $\omega_\pm(q)$ are obtained as
\begin{eqnarray}
\omega_{\pm}^2
=&&A^2+B^2-\tilde{V}_-^2\cos(\theta_1+\theta_2)
\nonumber \\
&&\pm\sqrt{\left[\tilde{V}_-\left(1-\cos(\theta_1+\theta_2)
\right)A\right]^2
+4B^2(A+C)(A-C)}\ ,
\label{equ:w13}
\end{eqnarray}
where $A={1\over2}(a+b)$, $B={1\over2}(a-b)$,
and $C={1\over2}\tilde{V}_-\left(1+\cos(\theta_1+\theta_2)\right)$, with
\begin{eqnarray}
a=&&\sqrt{(\Delta_{\rm sas}+\Delta_z+U_+\cos\theta_1
-U_-\cos\theta_2)^2+
(U_+\sin\theta_1+U_-\sin\theta_2)^2}-\tilde{V}_+, \nonumber \\
b=&&\sqrt{(\Delta_{\rm sas}-\Delta_z+U_+\cos\theta_2
-U_-\cos\theta_1)^2+
(U_+\sin\theta_2+U_-\sin\theta_1)^2}-\tilde{V}_+.
\label{equ:w12}
\end{eqnarray}

The intersubband SDW dispersions in both the canted antiferromagnetic
QH phase (C) and the normal QH phases (FM or SYM) can be obtained from the
above expression by incorporating appropriate values of $\theta_1$
and $\theta_2$.
In the following, we show $\omega_\pm(q)$
only at zero temperature for the sake of simplicity,
although the
formalism applies equally at finite temperatures.

In Fig. \ref{fig5}, we show the dispersion of the intersubband SDW
above the FM ground state.
As mentioned earlier, these two intersubband SDW modes
$\omega_{\pm}(q)$ correspond
respectively to transitions $|0\uparrow\rangle\rightarrow
|1\downarrow\rangle$ and $|1\uparrow\rangle\rightarrow
|0\downarrow\rangle$.
The frequencies $\omega_{\pm}$ increase as functions of $q$,
approaching asymptotic values $\omega_{\pm}(q\rightarrow\infty)
=\omega_{\pm}^0+|v_x|$,
where $\omega_{\pm}^0$
are the non-interacting excitation energies
and $v_x$ is the exchange energy of the electron
in the ground state.
Mode coupling, which
pushes down $\omega_-(q)$ and hence
helps mode softening,
is most visible at $q\rightarrow0$.
At zero layer separation, mode-coupling disappears,
and we recover previously known results.\cite{kal1,nu1}
In Fig. \ref{fig6}, we show the intersubband SDW dispersion
above the SYM state. The results are qualitatively
similar to
those in Fig. \ref{fig5}, except that there is no mode coupling
in the symmetric state because Coulomb interaction
is spin independent.
The important thing to be noticed is that the
long wavelength collective excitations
are gapped in both the symmetric phase and the spin polarized phase.
However, the mode softening does occur at the phase boundaries,
as we show below.

To illustrate the phase instability, we show, in Fig.\ref{fig8},
the lower-energy branch of the intersubband SDWs at
$q=0$ as a function of interlayer tunneling.
We see that $\omega_-(q=0)$ indeed softens
when approaching the phase boundaries
from both the symmetric phase and the spin polarized phase,
and remains zero inside the canted antiferromagnetic phase.
The canted antiferromagnetic order parameter, calculated
in the previous section, is also shown in Fig.\ref{fig8} for comparison
purpose. We notice that the phase boundaries
determined from these two independent approaches agree completely,
as shown in the figure.
The softening of the collective mode and
the appearance of the antiferromagnetic order parameter
implies that we have discovered
a quantum phase transition in double-layer QH systems.

In Fig.\ref{fig7}, the collective intersubband SDW dispersions
in the canted antiferromagnetic QH state are shown. The first
thing to be noticed is that the lower energy branch $\omega_-(q)$
is a gapless mode. The existence of such a
gapless Goldstone mode is due directly to
the canted antiferromagnetic
spin ordering which spontaneously breaks the spin-rotational
symmetry of the Hamiltonian.
This Goldstone mode is found to be
linear in the long wavelength limit, consistent
with the fact that it describes antiferromagnetic fluctuations.
The existence of the gapless excitation
in the canted antiferromagnetic phase implies
that some thermodynamic quantities, such as specific heat,
have power-law temperature dependence
in the canted antiferromagnetic phase
in contrast to their exponential temperature dependence in the normal
(symmetric or ferromagnetic) phases.

Simple expressions governing the phase boundaries
can be derived from the mode softening.
The boundary between the symmetric (SYM) phase and the
canted antiferromagnetic (C) phase is found to satisfy
the following equation
\begin{equation}
(\Delta_{\rm sas}-U_-)^2=U_-^2+\Delta_z^2,
\label{equ:bd1}
\end{equation}
where $U_-=\tilde{V}_-(q=0)
={1\over\Omega}\sum_{\bf p}e^{-p^2l_o^2/2}V_-(p)$.
It should be noted that, for any given $\Delta_{\rm sas}$,
the critical layer separation at this boundary is considerably smaller
than the critical layer separation
where the charge density excitation in the $\nu=1$ state
becomes soft.\cite{nu1}
The reason for this is the absence of Hartree contribution to
the SDW excitations.
The boundary between the spin polarized (FM) phase
and the canted antiferromagnetic (C) phase is found to satisfy
\begin{equation}
(\Delta_z+U_-)^2=U_-^2+\Delta_{\rm sas}^2.
\label{equ:bd2}
\end{equation}
The simplicity of Eqns. (\ref{equ:bd1}) and (\ref{equ:bd2})
makes the phase diagram easy to construct.
It is worthwhile to note that the phase boundaries are
determined by only three energy scales $\Delta_z$,
$\Delta_{\rm sas}$, and $U_-$
in spite of the fact that the Hamiltonian is determined by four
independent energy scales $\Delta_z$, $\Delta_{\rm sas}$,
and $V_\pm(q)$, of which
the inter- and intra-layer interactions $V_\pm(q)$ are in fact continuous
functions of wavelength $q$. This
unexpected dependence of the phase diagram (Fig. \ref{fig21})
on just three energy scales which are entirely determined by the
magnetic field, the sample parameters (i.e. inter-layer separation,
well width, etc.), and the tunneling strength, is a specific
result of the Hartree-Fock approximation.
The zero temperature phase diagram can thus be expressed
as a function of two independent dimensionless
variables $\Delta_z/\Delta_{\rm sas}$
and $U_-/\Delta_{\rm sas}$, as shown in Fig. \ref{fig21}.
This phase diagram applies to all double-layer quantum Hall
systems at $\nu=2$
which may have any values of Zeeman energy, tunneling energy,
layer separation, layer-thickness, etc.
We believe, however, that this remarkable scaling in the phase
diagram (which enables us to reduce an infinite number of
$\Delta_{sas}$ versus $d$ diagrams for various values of
$\Delta_z$, of which examples are shown in Fig. \ref{fig3},
to just one phase diagram shown in Fig. \ref{fig21})
remains approximately valid, although the relative size
of various phases in the universal
phase diagram of Fig. \ref{fig21} may very
well be quantitatively not particularly accurate.
 We also mention here that this
phase diagram  is
topologically identical to that of a $(2+1)$--dimensional
quantum $O(3)$
nonlinear $\sigma$-model in a magnetic field,\cite{dsz1}
as discussed in section \ref{nls} of this paper.

In this subsection, we have studied the collective intersubband
SDW excitations for $\nu=2$ double-layer QH systems
in the time dependent Hartree-Fock approximation.
We have presented the dispersions of the collective SDW excitations
in both the normal QH phases (FM and SYM) and in the canted antiferromagnetic
QH phase, and investigated the mode softening
which signals the phase instabilities.
We have rederived the same phase diagram as that obtained
in the previous section, and obtained analytic equations for the two phase
boundaries separating the new canted antiferromagnetic phase from
the normal FM and SYM phases.

\subsection{Kosterlitz-Thouless transition}
\label{KT}
In this subsection, we discuss some thermodynamic properties
of $\nu=2$ double-layer systems which arise from the spontaneous
symmetry-breaking
associated with the breaking of $U(1)$
planar spin rotational symmetry
in the canted antiferromagnetic quantum Hall phase.
There should be a finite temperature Kosterlitz-Thouless transition
in the canted antiferromagnetic phase, since the
spin-rotational
$U(1)$ symmetry is broken.
Below the critical temperature, the system supports
a linear Goldstone mode, which gives rise to a power-law
temperature dependence for the specific heat.
Above the critical temperature the $U(1)$ symmetry is restored and the
system is paramagnetic.
These properties are, in principle, experimentally observable
and provide direct ways to test our theory.

We can estimate the Kosterlitz-Thouless transition
temperature for our problem in the following manner.
In the canted
antiferromagnetic phase, the low temperature thermodynamics
is governed
by long wavelength phase fluctuations of
the order parameter.
Let $E_\phi=\langle\phi|H|\phi\rangle-\langle|H|\rangle$,
where $|\rangle$ is the ground state of the canted antiferromagnetic phase,
and $|\phi\rangle=\exp(i\sum_j S^z_j\phi_j)|\rangle$,
with $S^z_j$ as the spin operator of the j-th electron
and $\hat{z}$ is the (magnetic field)
direction normal to the two dimensional plane.
In the long wavelength limit, one obtains
\begin{equation}
E_\phi={\rho_s ( \Delta_z )\over2}\int d^2r |\bigtriangledown\phi({\bf r})|^2,
\label{equ:e1}
\end{equation}
with
\begin{equation}
\rho_s ( \Delta_z )={l_o^2\over16\pi^2}\int_0^\infty q^3 e^{-l_o^2q^2/2}
\left[v_a(q)\left({\sin{\theta_1}+\sin{\theta_2}\over2}\right)^2
+v_e(q)\left({\sin{\theta_1}-\sin{\theta_2}\over2}\right)^2\right]d q,
\label{equ:j1}
\end{equation}
where $l_o$ is the magnetic length
and $v_a$ ($v_e$) is intralayer (interlayer) Coulomb potential.
For future convenience, we have written the stiffness as an
explicit function of $\Delta_z$, which arises from the dependence
of the angles $\theta_{1,2}$ on the Zeeman splitting.
The effective planar $XY$\ model defined by Eq. (\ref{equ:e1}
undergoes a Kosterlitz-Thouless
phase transition \cite{kos1}
at approximately $k_BT_c=(\pi/2)\rho_s ( \Delta_z )$.
Finite temperature spin-wave and vortex-antivortex
polarizations reduce the transition temperature to
approximately
\cite{girvin1,girvin2,kos2}
\begin{equation}
k_BT_c\approx0.90\rho_s ( \Delta_z ).
\label{equ:t1}
\end{equation}
These finite temperature renormalizations can be much larger
in the vicinity of the C--N and C--SYM phase boundaries: the expression
(\ref{equ:t1}) can then no longer be used, and we will discuss modifications
near these boundaries later in Section~\ref{nls}.

Since we know $\rho_s ( \Delta_z )$ exactly within the microscopic
Hartree-Fock approximation,
the Kosterlitz-Thouless transition temperature can be easily determined for
our problem.
In Fig. \ref{fig33}, we show the calculated Kosterlitz-Thouless
critical temperature in $\nu=2$ double-layer quantum Hall systems
within the mean-field Hartree-Fock approximation (i.e. from
Eq. (\ref{equ:j1}) for $\rho_s ( \Delta_z )$).
The phase transition exists only in the canted antiferromagnetic quantum Hall
phase. The critical temperature vanishes at the phase boundaries
as the symmetry-breaking order parameter drops continuously to
zero as the phase boundaries are approached from within the canted
antiferromagnetic phase.
We notice that the calculated Kosterlitz-Thouless temperature ($\sim1K$)
is well within the experimentally accessible regime for
typical AlGaAs/GaAs--based double-layer systems.
The effective spin-stiffness $\rho_s ( \Delta_z )$ given in Eq. (\ref{equ:j1})
is obtained in
the mean-field Hartree-Fock approximation,
i.e. using the results from sections \ref{grou} and \ref{sdw},
where quantum fluctuation effects
are not included. The results
in Fig. \ref{fig33} should thus be regarded as the upper bound
for the Kosterlitz-Thouless critical temperature.
We emphasize that the Kosterlitz-Thouless transition discussed here
is present even in the presence of interlayer tunneling
(in fact, the presence of finite interlayer tunneling is essential to
stabilize the canted antiferromagnetic phase,
as described in the last two sections),
unlike the case associated with the pseudospin transition \cite{girvin1,girvin2}
at $\nu=1/m$ ($m$ odd integers) where interlayer tunneling
suppresses Kosterlitz-Thouless transition.

Below the Kosterlitz-Thouless
transition temperature, the specific heat
in the antiferromagnetic phase
has qualitatively different temperature dependence
from those of the normal quantum Hall phases.
This is of practical significance since it is possible
to experimentally measure the specific heat of
quantum Hall systems.\cite{sh1}
At low temperatures, the main contribution to
the specific heat comes
from long wavelength low energy
intersubband SDWs. With their dispersions calculated
in each of the quantum Hall phases, the heat capacity is
easily obtained:
$C=(\partial/\partial T)\sum_{\bf k}
\langle n_-(k) \rangle \omega_-(k)$,
where $\omega_-(k)$ is the energy of the low-lying intersubband
SDW excitation and $\langle n_-(k) \rangle$
is its Bose occupation factor.
The results are shown in Fig. \ref{fig34}.
It is clear that the specific heat has an activated behavior in the
normal quantum Hall phases because of the existence of excitation gap
in its spin wave spectra,
and a quadratic power-law temperature
dependence in the
canted antiferromagnetic phase
because of the existence of the linear Goldstone mode
in the symmetry broken phase.
The spin stiffness goes to zero discontinuously at $T_c$,
and for $T>T_c$ we have the usual disordered $X$--$Y$ phase
of the Kosterlitz-Thouless transition.

\subsection{Multicritical points}
\label{multi}

Our analysis so far has obtained solutions for the
FM, SYM and C phases obtained by varying the parameters
$\Delta_z$, $\Delta_{\rm sas}$, $d$ in the Hamiltonian
(see Figs. \ref{fig3} and \ref{fig21}) which modify
the relative strengths of the Zeeman energy, the tunneling energy,
and the Coulomb interaction energy, respectively.
Generically, these phases separated by phase boundaries
representing second-order quantum transitions. However,
there are also special quantum  
multicritical points in Figs~\ref{fig3}
and~\ref{fig21} whose physical significance we will now discuss.

The first quantum multicritical point is apparent in Fig. \ref{fig3}
where the FM, C and SYM phases come together
at a single point on the abscissa corresponding to vanishing interlayer
separation ($d=0$).  It is easily noted from Fig. \ref{fig3} that
this quantum multicritical point is in fact defined by
\begin{equation}
\Delta_{\rm sas}=\Delta_z;\ \ \ \ d=0,
\label{equ:sds1}
\end{equation}
which is equivalent to the conditions
\begin{equation}
\Delta_{\rm sas}=\Delta_z;\ \ \ \ V_-(q)=0,
\label{equ:sds2}
\end{equation}
using the definition of $V_\pm(q)$ given immediately
following Eq. (\ref{equ:hi}) in section \ref{grou}.
The simple physical reason for the vanishing of $V_-(q)$ along the
$d=0$ line is that the intra- and interlayer Coulomb interactions
are identical in the limit of vanishing interlayer separation $d$.
Note also that the vanishing of $V_-$ ( and consequently
of $U_-$) pushes the quantum multicritical point to an infinite value
of the abscissa ( $\Delta_{\rm sas}/U_-\rightarrow\infty$) in the
scaled universal diagram given in Fig. \ref{fig21}---
in Fig. \ref{fig21} the two phase boundaries separating the
three phases approach each other asymptotically
as $\Delta_{\rm sas}/U_-\rightarrow\infty$
and $\Delta_z/\Delta_{\rm sas}\rightarrow1$ at the
multicritical point.
Note that the condition $\Delta_z=\Delta_{\rm sas}$
for the quantum multicritical point is a particularly
interesting criterion because, in the absence of
our predicted canted antiferromagnetic phase (i.e. if
the $\nu=2$ double-layer QH systems allowed only
the ferromagnetic and the symmetric phases, as was assumed
in the literature before our work),
the condition of the equality of the Zeeman splitting
and the symmetric-antisymmetric gap (i.e. $\Delta_z=\Delta_{\rm sas}$)
is precisely the single particle {\sl level crossing}
criterion where, at $\nu=2$, one would make a transition from the ferromagnetic
phases where the two up-spin symmetric and antisymmetric
levels are occupied and the down-spin levels are empty
for $\Delta_z>\Delta_{\rm sas}$ to the
symmetric (spin singlet) phase where the spin-up
and spin-down symmetric subbands are occupied (and the antisymmetric levels
are empty ) for $\Delta_{\rm sas}>\Delta_z$.
What our theory definitely predicts is that such a simplistic
one particle level crossing picture
(which appears to be obvious intuitively) {\sl does not occur}
in a double-layer QH system at $\nu=2$---
instead Coulomb interaction breaks the $SU(2)$ spin rotational symmetry
and drives the system into an intervening antiferromagnetic phase
where spin and pseudospin levels are intrinsically mixed.
The fact that the intuitively expected level crossing
phenomenon (at $\Delta_z=\Delta_{\rm sas}$) has never been observed
\cite{pc1}
in a $\nu=2$ double-layer QH system in spite of
systematic efforts \cite{pc2}
is, in our opinion, rather strong indirect evidence in
support of our phase diagram.

The second multicritical point becomes apparent only in the universal
phase diagram shown in Fig. \ref{fig21} (and can be inferred implicitly
from the trend that can be seen in the phase diagrams shown in
Fig. \ref{fig3}). Its existence is a consequence of the intriguing finding
that our
antiferromagnetic state, in fact, persists all the way to
$\Delta_z=0$ (as can be clearly seen in Fig. \ref{fig21}
where a finite region of the antiferromagnetic state exists along
the $\Delta_z=0$ line) where the spin-polarized ferromagnetic
phase no longer exists, and the
antiferromagnetic phase is separated from the spin singlet phase by a
multicritical point (M) defined by the condition
\begin{equation}
\Delta_{\rm sas}=2U_-\ \ \ \ {\rm with}\ \ \ \Delta_z=0.
\label{equ:sds3}
\end{equation}
Thus the critical line defining the phase boundary between the
antiferromagnetic and the symmetric phases
for $\Delta_z\ne0$ ends at a critical point (M)
for $\Delta_z=0$.
It is evident that in the absence of any Zeeman energy ($\Delta_z=0$)
the spin magnetic moment in each layer lies completely in the 2D plane of the
electron gas where they must be equal and opposite in the two layers.
Therefore, the $\Delta_z=0$ antiferromagnetic phase
of Fig. \ref{fig21} is {\sl not} a {\sl canted} phase,
but is a purely N\'{e}el phase (N); indeed the Hamiltonian has
full $SU(2)$ spin rotation symmetry for $\Delta_z = 0$, and spin
moments in the N phase can point in any two anti-parallel
directions. The N, C, and SYM phases meet at the multicritical
point M. This multicritical point will take on
special significance in our effective field theoretical formulation
in the next section.

Let us also note that
the existence of this purely N\'{e}el QH antiferromagnet
at $\nu=2$ double-layer system may not be just a theoretical curiosity
because it is possible to obtain vanishing Zeeman splitting
in a GaAs double-layer system in a finite magnetic field situation
by applying external pressure which under suitable conditions could
lead to the vanishing of the effective gyromagnetic ratio
(the $g$-factor)
due to band structure effects.

\subsection{Comparison with earlier work}
\label{earlier}

Before concluding this section, and going on to the effective field theoretic
description of the double-layer QH system, we will discuss the
relationship of our results to some earlier work on double layer
systems. We will also use this opportunity to comment on the
validity of the Hartree Fock approximation in our and earlier
work.

Most earlier studies,\cite{kun1,girvin1,girvin2,nu1} however, have focused on
$\nu=1$ (with some work\cite{nuh} on $\nu=1/2$).
Although the $\nu=1$ and the
$\nu=2$ QH systems exhibit some similarities such as the softening
of their low energy collective excitations under certain conditions,
there are important distinctions between them.
At $\nu=1$, the spin degree of freedom is normally frozen out
by the external magnetic field.
The relevant low energy excitations in the $\nu=1$ QH state are
therefore intersubband charge-density-wave excitations,
and the properties of the system are determined by the interplay between
interlayer tunneling energy and Coulomb interaction energy.
In this sense, the $\nu=1$ system is in fact a single-layer system with
a layer pseudospin-dependent interaction.\cite{girvin1,girvin2,nu1}
At $\nu=2$,
both the spin degree of
freedom and the layer degree of freedom are relevant,
and the low energy excitations are intersubband SDW excitations.
Consequently, the properties of the system are determined
by the interplay among tunneling energy, Zeeman energy,
and Coulomb interaction energy.
Because of the increased degree of freedom, the system has more ways
to optimize the total energy, and new states which are not
possible at $\nu=1$ become possible at $\nu=2$.
The symmetric QH state is energetically favored at
small layer separations because it optimizes the tunneling energy;
The spin polarized QH state is favored at large layer separations
because it optimizes the Coulomb interaction energy;
The canted antiferromagnetic state
is energetically favored at intermediate layer separations.
The reason for this
is that the canted antiferromagnetic state tends to simultaneously
optimize both the tunneling
energy and the Coulomb interaction energy, which prevails
at intermediate layer separations where
the tunneling energy and the Coulomb interaction energy
are equally important.
Both the canted antiferromagnetic state and the symmetric state
exist only for systems with small enough
Zeeman energy, as the Zeeman energy clearly favors the
spin polarized state.

Another important distinction between the $\nu=1$ systems and
the $\nu=2$ systems
is that although
at $\nu=1$ the mode softening destroys the QH effect\cite{nu1},
and there is no reliable description of the electronic state
in the non-QH phase because beyond the critical layer separation
the system becomes effectively a pair of
isolated layers with compressible half-filled Landau level states;
in contrast, at $\nu=2$, the QH effect prevails at all phases
because there is always a charge gap in both layers (even as $d \rightarrow
\infty$),
and we have good understanding of the ground state and the
low energy excitations in each phase
due to the existence of incompressible filled Landau levels.
Nevertheless, the mode softening and the associated
phase transitions at $\nu=2$ are likely to be
observable through inelastic
light scattering experiments\cite{ap1,ap2}
and thermodynamic measurements.

Our work has studied
$\nu=2$ double-layer systems by numerically solving
the self-consistent mean-field
equations\cite{wig1}, and obtained collective
excitation dispersions using many-body diagrammatic techniques.
\cite{kal1}
Both approaches are, however, based on the Hartree-Fock
approximation.
In single-layer integer QH systems,
calculations \cite{kal1} in the Hartree-Fock approximation
agree well with experiments.\cite{ap2}
In double-layer systems, the Hartree-Fock approximation
is less accurate because Coulomb interaction potential is
layer-index dependent. Nevertheless, we expect
that the Hartree-Fock approximation
remains a reasonably good description for a double-layer system at $\nu=2$,
since the Hartree-Fock ground state,
which is non-degenerate and separated in energy from higher levels,
is a good approximation for the real many-body ground state
at $\nu=2$
due to the existence of
incompressible filled Landau level states
with charge excitation gaps at any layer separations.
We want to especially emphasize the difference in the validity of the
Hartree-Fock approximation between $\nu=1$ and $\nu=2$.
The approximation is valid at $\nu=1$ only at small layer separations
and fails completely beyond a critical layer separation where
the system becomes effectively a pair of isolated
layers with compressible half-filled Landau level state in each layer.
At $\nu=2$, incompressible states with filled Landau levels
exist at any layer separations. In particular,
there is still one filled Landau level in each layer
at $d\rightarrow\infty$.
This fact, namely the existence of an incompressible
energy gap at all layer separations,
ensures that the Hartree-Fock approximation,
upon which our calculations are based,
is a reasonable formalism at $\nu=2$
regardless of the value of the layer separation.

\section{Continuum field theory and quantum critical phenomena}
\label{nls}

The Hartree-Fock analysis used in the previous sections has the advantage
of working with a realistic microscopic Hamiltonian and of making definite
quantitative predictions for experimental observables
in realistic samples.
In this section, we will present an alternative analysis based upon
a continuum effective
quantum field theory for the low-lying spin excitations of
a double layer quantum Hall system.
We will find that the global phase diagrams obtained in the two approaches
are very similar,
and are, in fact, topologically identical,
and that detailed additional predictions for the
temperature dependence of various observables can be made by a combination
of the two methods. In particular, some advantages offered by the
continuum approach are:
\begin{itemize}
\item
It will become clear from the analysis below that there are
two basic ingredients necessary to obtain the phases of Fig~\ref{fig21}:
two well separated layers form fully polarized ferromagnets with a gap
towards charged excitations
(i.e. an incompressible QH effect gap), and the primary coupling between
the layers
is an antiferromagnetic exchange
(i.e. a superexchange) interaction. As such, we expect
a similar phase diagram to apply not only at filling $\nu = 2$,
but also at any $\nu = 2 \nu_1$, where $\nu_1$ is any filling fraction
where a single layer has a charge gap, and is fully polarized.
In particular, this criterion is satisfied at $\nu =2/m$, $m$ an odd integer,
where each layer forms a polarized Laughlin fractional quantum Hall state.
The Hartree-Fock analysis clearly cannot be applied for $m>1$,
as the single layer charge gap appears only after inclusion of the non-trivial
correlations implicit in the Laughlin state.
\item
The Hartree-Fock theory significantly overestimates the energy of the
spin-unpolarized
symmetric (SYM) or the spin singlet (SS)
state, as we will refer to it in this section.
Spin up and spin down electrons are simply placed
into the same orbitals which are symmetric in the layer index. This is
costly in Coulomb energy as there are no correlations in the layer positions
of the up and down spin electrons. It is clearly more advantageous to
form a spin singlet states between electrons which are localized in opposite
layers. The non-linear sigma model continuum field theory to be discussed
below does this in a natural way. From now on in this section
we refer to this symmetric or the
spin singlet phase as the SS to emphasize its correlated singlet nature.
\item
A number of quantum-critical points have been uncovered in the Hartree-Fock
analysis. There is the $\Delta_z =0$ quantum critical point between the
spin singlet (SS or SYM)
and the N\'{e}el (N)
phases, and a critical line between the SS
and the canted antiferromagnetic (C) phases. Our continuum
approach will obtain the critical theory for these transitions,
and we will find that they have dynamic critical exponents
\cite{son1} $z=1$ and $z=2$
respectively. There is also a second critical line between the C and
the fully spin polarized ferromagnetic (FM)
phases: this transition has $z=2$ and will be discussed only in passing,
as the critical theory is rather similar to one of the models discussed
in detail in Ref~\onlinecite{annals}.
\item
The continuum theory offers
not only provides us the zero temperature quantum phase diagram but also
 a streamlined approach to the study of
properties at non-zero temperature, especially in the vicinity of the
quantum critical points where effects of fluctuations cannot be neglected.
The price one pays is that in general the parameters defining the
effective field theory are quantitatively unknown
and can only be calculated from a microscopic theory such as
the Hartree-Fock theory of the previous sections.
\end{itemize}

We motivate our formulation of the continuum theory by consideration
of the physics of two well-separated identical layers at $\nu = 2/m$. More
specifically, the layer separation, $d$, is much larger
than the magnetic length, $\ell_o$. Then the two layers (labeled 1,2)
are approximately decoupled, and each separately has filling fraction
$\nu_1 = \nu_2 = 1/m$. Their ground states
will be
the familiar Laughlin states for $m>1$, or a fully filled lowest Landau
level
at $m=1$, both of which are incompressible states with
large energy gap to
all charged excitations.
These states are also fully spin polarized; the spin polarization is
induced not just by the Zeeman coupling to the external magnetic
field, but also by the significantly larger
intralayer ferromagnetic exchange~\cite{girvin1,girvin2,sondhi}.
The low-lying
excitations in each layer are spin waves which have a small excitation
gap given precisely
by the Zeeman energy $\Delta_z = g \mu_B B$.
For small $g$, a complete description~\cite{girvin2,nr} of the low
energy
excitations of each layer
can be given in terms of an action for unit vector fields
$\vec{n}_{1,2}$ (
$\vec{n}_{1,2}^2 = 1$) representing the orientation of the ferromagnetic
orders.
Spin waves are small fluctuations of $\vec{n}_{1,2}$ about an ordered
state,
while charged quasiparticles are Skyrmion~\cite{girvin2,leekane,sondhi,fertig}
textures of $\vec{n}_{1,2}$.
The effective action describing the two layers
is~\cite{girvin1,girvin2,sondhi,nr}
(in units with $\hbar = k_B = 1$)
\begin{eqnarray}
&& {\cal S}_0 = \int d^2 x \int_0^{1/T}  d \tau
\left( {\cal L}_F [\vec{n}_1] + {\cal L}_F [\vec{n}_2 ]
\right) \nonumber \\
&& {\cal L}_F[\vec{n}] \equiv iM_0 \vec{A}(\vec{n}) \cdot
\frac{\partial \vec{n}}{\partial \tau}
+ \frac{\rho_s^0}{2} \left( \nabla_x \vec{n} \right)^2 - M_0 \Delta_z
n_z
\label{action0}
\end{eqnarray}
Here
\begin{equation}
M_0 = \frac{1}{4 \pi m \ell_o^2}
\end{equation}
is the magnetization density per layer, with $l_o$ the magnetic length.
 The spin stiffness of each
well separated layer is represented by $\rho_s^0$;
for $m=1$, we have the exact result~\cite{kal1}
$\rho_s^0 = e^2/(16 \sqrt{2\pi} \epsilon \ell_o)$, while for $m>1$ numerical
estimates of $\rho_s^0$ are given in Ref~\cite{sondhi}.
The term involving $\vec{A}$ accounts
for the Berry phase accumulated under time evolution of the spins;
here $\vec{A}$ is any functional of $\vec{n}$ which satisfies
\begin{equation}
\epsilon_{ijk} \frac{\partial A_k (n)}{\partial n_j} = n_i.
\end{equation}
This Berry phase term also has a ``dual'' interpretation in the picture in which
${\cal L}_F$ is viewed as an action for Skyrmions~\cite{nr,stone}: it
represents the coupling
of the Skyrmion current to a ``magnetic field'' of strength $4 \pi M_0$.

Now imagine reducing the value of $d$ to couple the two layers.
As there is a charge gap in each layer, we can neglect all
charge transfer processes, and focus solely on spin exchange.
Because of the strong repulsive interactions within each layer, we expect
by an extension
of the familiar arguments made in the context of the Hubbard model that
there will be an {\em antiferromagnetic} superexchange coupling between the
layers.
This can also be inferred easily by considering the leading effect of
interlayer tunneling and Pauli principle, which immediately provides
a superexchange coupling between the layers.
The complete double layer action is therefore
\begin{equation}
{\cal S}_1 = \int d^2 x \int_0^{1/T}  d \tau
\left( {\cal L}_F [\vec{n}_1] + {\cal L}_F [\vec{n}_2 ] + J \vec{n}_1
\cdot \vec{n}_2
\right)
\label{action1}
\end{equation}
The value of the interlayer exchange, $J$, is not known
precisely; we expect that it is of order
$J \sim M_0
\Delta^2_{\rm sas}/ U$
where
$\Delta_{\rm sas}$
is the tunneling matrix element
(see Eq. (\ref{equ:h1}), for example) between the layers, and $U \sim e^2
/ \epsilon \ell_o$
is the Coulomb repulsion energy.
In addition to the imprecisely known $J$, the present approach also requires
knowledge of the nature of the short distance cutoff at lengths of order
$\ell_0$
beyond which present continuum approach cannot be applied.
We will show later that our ignorance of these quantities can be reduced
entirely to
uncertainties in the value of a certain renormalized energy scale. This energy
scale can be either measured directly in an experiment, or computed by
any microscopic theory such as the
Hartree-Fock approach (appropriate at $\nu=2$)
described in sections \ref{grou}--\ref{KT}.
Apart, from this single energy scale, however,
all of the predictions of the present
effective field theoretical approach will be quantitative and precise.

Some potentially important terms have been omitted from ${\cal S}_0$
and our analytic computations: the Hopf term which endows the Skyrmions
with fractional statistics,
and the long-range Coulomb
interaction between the Skyrmions. We believe this is permissible
because of the charge
gap. Further~\cite{scs}, as the layers are
antiferromagnetically correlated,
Skyrmions in one layer will be correlated with anti-Skyrmions in the
other, and this
neutralizes the leading contribution of both terms. This latter argument
should continue to hold
even if the charge gap were to vanish at a quantum critical point (the charge
gap remains non-zero at the quantum critical points in both our present
calculations).
Note also that no new term is necessary to induce charge transfer
between the layers:
a hedgehog/anti-hedgehog pair in the two layers corresponds to an event
transferring
Skyrmion number between them. Such spacetime singularities are absent in
the
strict continuum limit but appear when a short-distance regularization
is introduced.

For completeness, we note that the purely $\vec{n}$ field formulation
becomes incomplete for $m>1$ and larger $g$, as the spin zero Laughlin
quasiparticles can become the lowest energy charged excitations. These
should, in principle,
be accounted for by a separate complex scalar field. However, these can
also be neglected for the same reasons presented above for non-zero spin
charged excitations.

We now manipulate the effective action
into a form more suitable for our subsequent analysis.
We solve the constraints $\vec{n}_{1,2}^2 = 1$ by representing
\begin{equation}
\vec{n}_i = (-1)^i (1 - \vec{L}^2 )^{1/2} \vec{n} + \vec{L}
\label{nl}
\end{equation}
where $\vec{n}$ and $\vec{L}$ are vectors satisfying
\begin{equation}
\vec{n}^2 = 1~~~~~~~~~\vec{L} \cdot \vec{n} = 0.
\end{equation}
Note that this representation is so far exact. Next, we insert
(\ref{nl}) into ${\cal S}_1$.
Because the layers are
antiferromagnetically correlated
we expect that $\vec{L}$ will not be too large, and it is therefore
permissible to expand the resulting action to quadratic order in $\vec{L}$.
This is clearly an approximation: in Appendix~\ref{twospin} we examine
a model solvable Hamiltonian by the same method in order to assess the damage
done by neglecting terms higher order in $\vec{L}$---we find that this procedure
obtains the low energy spectrum correctly but introduces some spurious states
at higher energies.
To quadratic order in $\vec{L}$, ${\cal S}_1$ takes the form
\begin{equation}
{\cal S}_1 = \int d^2 x \int_0^{1/T} d \tau
\left[ 2 i M_0 \vec{L} \cdot \left( \vec{n} \times
\frac{\partial \vec{n}}{\partial \tau} + i \Delta_z \hat{z} \right)
+ \rho_s^0 \left( \nabla_x \vec{n} \right)^2 + 2 J \vec{L}^2
\right],
\label{action1a}
\end{equation}
where $\hat{z}$ is a unit vector in the direction of the magnetic field.
Now we integrate out $\vec{L}$ while maintaining the constraint
$\vec{L}\cdot \vec{n} = 0$ by adding an additional term to the energy
$\sim C ( \vec{L} \cdot \vec{n} )^2$ and then taking the limit $C
\rightarrow \infty$.
This yields the
following
effective action for the antiferromagnetic order parameter $\vec{n}$
\begin{equation}
{\cal S}_2 = \frac{c}{2t} \int d^2 x \int_0^{1/T}  d \tau
\left[ (\nabla_x \vec{n} )^2 + \frac{1}{c^2} \left( \frac{\partial
\vec{n}}{\partial \tau}
- i \Delta_z \hat{z} \times \vec{n} \right)^2 \right]
\label{action2}
\end{equation}
where
\begin{equation}
t = \left( \frac{J}{ 2 \rho_s^0 M_0^2}\right )^{1/2}~~~~~~~~
c= \left( \frac{2 \rho_s^0
J}{M_0^2} \right)^{1/2}.
\label{equ:tt1}
\end{equation}
This is precisely the action of the 2+1 dimensional quantum $O(3)$
non-linear
sigma model in a field $B$ coupling to the conserved global $O(3)$
charge.\cite{ss1,annals,ss3,ss2} It is expected
to apply to double-layer quantum Hall systems with $\nu = 2/m$
at length
scales larger
than $\Lambda^{-1} \sim \ell_o$.

The remainder of this section consists of a detailed analysis of the
properties of ${\cal S}_2$. The techniques and some results have already
been presented earlier in Refs~\onlinecite{ss1,annals,ss2}: we shall
present here
a unified treatment with a special emphasis on dynamical properties at
non-zero temperature. We begin in Section~\ref{mft} by developing a simple
mean-field phase diagram of ${\cal S}_2$.

\subsection{Mean field theory}
\label{mft}

This section will summarize the results of the application of the mean field
theory of Ref~\onlinecite{annals} to the action ${\cal S}_2$. Formulation of the
mean field theory requires some short distance regularization, and we
choose to place
the continuum theory on a square
lattice in the spatial directions, with a lattice spacing $a \sim \ell_o$;
a continuum 
formulation is maintained along the time direction. The resulting action is
equivalent
to the following lattice quantum rotor Hamiltonian
\begin{equation}
{\cal H} = \sum_i \left( \frac{f}{2} \hat{\vec{L}}_i^{2} - \Delta_z \hat{z}
\cdot \hat{\vec{L}}_i
\right) - K \sum_{<i,j>} \hat{\vec{n}}_i \cdot \hat{\vec{n}}_j
\label{hrotor}
\end{equation}
where the coupling constants in ${\cal H}$ are
\begin{equation}
f = \frac{c t}{a^2}~~~~~~~~~~~~~K = \frac{c}{t}
\end{equation}
The Hamiltonian is expressed in terms of operators $\hat{\vec{n}}_i$ which
represent the
orientation of the rotors, and $\hat{\vec{L}}_i$ which are the rotor
angular momenta.
The operators on different sites commute, while those on a single site obey the
commutation relations (dropping the site indices)
\begin{eqnarray}
\left[ \hat{L}_{\alpha}, \hat{L}_{\beta} \right] &=& i
\epsilon_{\alpha\beta\gamma}
\hat{L}_{\gamma} \nonumber \\
\left[ \hat{L}_{\alpha}, \hat{n}_{\beta} \right] &=& i
\epsilon_{\alpha\beta\gamma}
\hat{n}_{\gamma} \nonumber \\
\left[ \hat{n}_{\alpha}, \hat{n}_{\beta} \right] &=& 0
\end{eqnarray}

We will describe the properties of ${\cal H}$ by choosing the best among
the mean field
Hamiltonians given by~\cite{annals}
\begin{equation}
{\cal H}_{MF} = \sum_i \left( \frac{f}{2} \hat{\vec{L}}_i^{2} - \Delta_z \hat{z}
\cdot \hat{\vec{L}}_i
- K Z \vec{N} \cdot \hat{\vec{n}}_i\right)
\end{equation}
Here $Z$ ($=4$) is the lattice co-ordination number, and $\vec{N}$ is a
variational
parameter to be chosen so that the expectation value of ${\cal H}$ in the ground
state of ${\cal H}_{MF}$ is as low as possible; by the usual argument, this
is expected
to happen when $\vec{N} = \langle \hat{\vec{n}} \rangle$.

As in Ref~\onlinecite{annals}, we numerically diagonalized ${\cal H}_{MF}$
by truncating its spectrum at some large angular momentum, and then
optimized the
value of $\vec{N}$. The resulting phase diagram is shown in Fig~\ref{mftfig}.
We discuss the properties of the various phases in turn:

\subsubsection{Spin Singlet (SS or SYM)}
Each rotor is in its non-degenerate $\ell=0$ state, $\vec{N}=0$, and there
is a gap
to all excitations. The ground state is a spin singlet, and is therefore
unaffected
by variations in the value of $\Delta_z$.

\subsubsection{Quantized Ferromagnets (QF${}_{\ell}$)}
Again, $\vec{N}=0$,
each rotor now has azimuthal angular momentum $m=\ell$ and this value remains
pinned as various parameters are varied. Each rotor is also in precisely
the state with $\hat{\vec{L}}^2 = \ell ( \ell + 1)$, although this latter
feature
is a special
property of mean field theory which will not survive fluctuations.
Of these phases, only the $\ell =1$ case is actually allowed for the double
layer
action ${\cal S}_1$, and it is clearly the FM phase of Fig~\ref{fig21}.
The other phases are an artifact of the approximations made
in mapping ${\cal S}_1$ to ${\cal S}_2$: this should be clear from the
discussion
in Appendix~\ref{twospin} where we show that expanding in powers of
$\vec{L}$ introduce
spurious higher angular momenta states.

\subsubsection{Canted (C) and N\'{e}el (N) States}
These states have $\vec{N} \neq 0$ and varying continuously as the
parameters are
varied; we have $N_x \neq 0$, $N_y \neq 0$
and $N_z = 0$.  From (\ref{nl}), this implies that the two layers have opposite
spin polarizations in the $x-y$ plane. The two layers also have an identical
ferromagnetic polarization, given by $\langle \hat{\vec{L}} \rangle$ which
is oriented
along the $z$ direction. This ferromagnetic moment varies continuously as
parameters
are varied, and vanishes when $\Delta_z \rightarrow 0$. So for general
$\Delta_z \neq 0$
 this state is canted (C), while for $\Delta_z=0$ it is a pure N\'{e}el (N)
antiferromagnet.
The C phase has a single linearly dispersing spin wave mode
in the $x-y$ plane, while the N phase has two spin waves~\cite{annals}.

In the remainder of this section, we will present a detailed theory of the
universal
properties of the system in the vicinity of the multicritical point M.
This is the same quantum multicritical point (M) which
exists in the universal Hartree-Fock phase diagram of Fig. \ref{fig21}
where the N phase (along the $\Delta_z=0$ line),
the C phase and the SYM (SS) phase come together at $\Delta_{\rm sas}=2U_-$.
We point out in this context that the other distinct multicritical point
of the Hartree-Fock theory where the canted antiferromagnetic phase,
the ferromagnetic phase, and the symmetric phase coexist
(the point on the abscissa defined by $d=0$ and $\Delta_z=\Delta_{\rm sas}$
in Fig. \ref{fig3}) is not accessible within the effective field theory
due to the long wavelength restriction $d>l_o$.
(We mention that in our notations $\Delta_z$ in the Hartree-Fock theory
corresponds to just $B$ in the field theory due to
our choice of units.)

Note that
the C, N, and SS phases meet at M, and so we will also discuss the universal
second-order transitions between them. We will not discuss the nature of
the second-order
transitions between the $QF_{\ell}$ and $C$ phases: very closely related
transitions,
in the same universality class, have been discussed in some detail in
Ref~\onlinecite{annals}.

\subsection{Zero temperature critical properties}

A first study of the properties in the vicinity of the point M has appeared
in Ref~\onlinecite{ss1} using a large $N$ expansion in a non-linear sigma
model with $N$ component fields. The issues of interest here are more
conveniently obtained
using a recently developed expansion~\cite{ss2} in spatial dimensionality,
$d$, in powers of $\epsilon=3-d$.
The latter approach is expressed in terms of a soft-spin field theory,
and we therefore begin with a soft-spin version of the non-linear sigma model
${\cal S}_2$:
\begin{eqnarray}
{\cal S}_{\phi} = \int d^d x \int_0^{1/T} && \left[
\frac{1}{2} \left\{
( \partial_{\tau} \phi_x - i \Delta_z \phi_y )^2 +
( \partial_{\tau} \phi_y + i \Delta_z \phi_x )^2 +
( \partial_{\tau} \phi_z )^2 + c^2 (\nabla_x \vec{\phi})^2 \right. \right.
\nonumber \\
&& ~~~~~~~~~~\left. \left. + r \vec{\phi}^2 \right\}
+ \frac{u_0}{4!} ( \vec{\phi}^2 )^2 \right]
\label{hphi}
\end{eqnarray}
Here $\vec{\phi} \equiv ( \phi_x, \phi_y, \phi_z ) \sim \vec{n}$ is the
soft-spin field
which measures the staggered moment of the two layers. We have taken the
magnetic field to point in the $z$ direction. We will also be interested in the
uniform ferromagnetic moment density of the system, $M$, and this is given by
\begin{equation}
M \equiv M_0 \langle {n}_{1z} + {n}_{2z} \rangle = - \frac{\partial
{\cal F}}{\partial \Delta_z}
\end{equation}
where ${\cal F}$ is the free energy density associated with the action
${\cal S}_{\phi}$.
We have introduced two new coupling constants, $r$ and $u_0$ in ${\cal
S}_{\phi}$;
these are related to the coupling $t$ of ${\cal S}_2$,
and its short-cutoff $\sim \ell_o$. We will not specify the precise values
of these
parameters here, as they merely appear at intermediate stages of our
computation,
and not in our final results.

Let us first discuss the mean field properties of ${\cal S}_{\phi}$,
obtained simply
by minimizing the action while ignoring all spatial and time dependence of
$\vec{\phi}$.
For $r - \Delta_z^2 > 0$, the ground state has $\langle \vec{\phi} \rangle
= 0$, and is
therefore in the quantum paramagnetic SS phase. For $r - \Delta_z^2 < 0$,
the ground state
has $\langle \vec{\phi} \rangle \neq 0$ and in the $x-y$ plane. This is the
C phase and the fields have the expectation values
\begin{equation}
\vec{\phi} = \left( \left( \frac{6 (\Delta_z^2-r)}{u_0} \right)^{1/2}, 0,
0\right)~~~~~~~~
M = \frac{6\Delta_z (\Delta_z^2 - r)}{u_0},
\end{equation}
or any rotation of $\vec{\phi}$ in the $x-y$ plane. Notice that $M$
vanishes for $\Delta_z=0$,
and therefore the line $r<0$, $\Delta_z=0$ is the N phase. The resulting
mean field phase
diagram is shown in Fig~\ref{landau}. Notice that the vicinities of the
points M are very
similar in Figs~\ref{mftfig} and~\ref{landau}. The quantum critical point M is
at $\Delta_z=0$, $r=0$, and it is clear from the Lorentz-invariant
structure of ${\cal S}_{\phi}$ at $\Delta_z =0$ that this point has dynamic
exponent
$z=1$. Rotations of the order parameter $\vec{\phi}$ in the $x-y$ plane
have associated
with them a stiffness $\rho_s ( \Delta_z)$ given by
\begin{equation}
\rho_s ( \Delta_z ) = \frac{6 ( \Delta_z^2 - r)}{u_0}.
\label{rhosmft}
\end{equation}
This is the same stiffness which was computed in Section~\ref{KT} and
Eqn~\ref{equ:j1} in the Hartree-Fock theory.

We now include the effects of fluctuations at one loop. We will quote
results for the
dynamic longitudinal and transverse susceptibilities of the $\vec{\phi}$
field which
are measured in light scattering.
Recall that in terms of the spin polarizations of the two layers
$\vec{n}_1$, $\vec{n}_2$,
we have $\vec{\phi} \sim \vec{n}_1 - \vec{n}_2$. We define
(with $T$ as the temperature $k_B=1$ in our units in this section)
\begin{equation}
\chi_{\parallel} ( i \omega ) = \int d^d x \int_0^{1/T} d \tau e^{-i \omega
\tau}
\left\langle \phi_z (x, \tau) \phi_z (0,0) \right\rangle
\end{equation}
and
\begin{equation}
\chi_{\perp} ( i \omega ) = \frac{1}{2} \int d^d x \int_0^{1/T} d \tau
e^{-i \omega \tau}
\left\langle (\phi_x (x, \tau) + i \phi_y (x, \tau))( \phi_x (0,0)
-i \phi_y (0,0)) \right\rangle
\end{equation}
We can use the methods of Ref~\onlinecite{ss2} to compute the one loop
values of these susceptibilities in the SS phase (this is the phase with no
broken symmetry)
in the vicinity of the
point M; we obtain
\begin{eqnarray}
\chi_{\parallel} ( \omega ) &=& \frac{1}{\Delta^2 - \omega^2} \nonumber \\
\chi_{\perp} ( \omega ) &=& \frac{1}{\Delta^2 - (\omega-\Delta_z)^2}
\label{chiteq0}
\end{eqnarray}
Here the quantity $\Delta$ is an observable defined by
\begin{equation}
\Delta \equiv \mbox{Spin gap of the SS phase at $T=0$ for $r>0$ and
$\Delta_z =0$}.
\end{equation}
The value of $\Delta$ should either be measured experimentally, or computed by
a detailed microscopic calculation like the Hartree-Fock theory discussed
earlier in the
paper. We will express all our results for $r>0$ completely in terms of
universal functions
of parameters
$\Delta$, $T$ and $B$ (so that
the microscopic couplings $r$ and $u_0$ do  not appear
anywhere in our results.) Clearly, in the
mean field theory $\Delta = \sqrt{r}$; at one loop
order, we have $\Delta \sim r^{\nu}$, where the exponent $\nu = 1/2 + 5
\epsilon/44$.

We need a separate experimental observable to measure the deviation of the
system
from the point $M$ at $\Delta_z =0$ for $r<0$. A convenient choice, also
used in 
Refs~\onlinecite{ss3,ss2} is the spin stiffness. We therefore define
\begin{equation}
\rho_s (0) \equiv \mbox{Renormalized spin stiffness of the N phase at $T=0$ for
$r<0$ and $\Delta_z=0$}.
\end{equation}
All our results for $r<0$ will be expressed in terms of $\rho_s (0)$.
Again $\rho_s (0) \sim |r|^{\nu}$, and the actual value of $\rho_s (0)$
should be
measured experimentally or computed in Hartree-Fock or microscopic
numerical studies
of the double-layer Hamiltonian.

Before closing this subsection, we draw attention to the fact that there are
two phase boundaries that terminate at the point M: the SS to C transition
and the N to C transition. In the vicinity of these transitions the response
functions computed near
the critical point M should turn into {\em reduced} scaling
functions~\cite{ss2,madrid}
characteristic of the respective phase transitions.  In the following
subsections,
we discuss simplified versions
of the action ${\cal S}_{\phi}$ which can be used to compute these reduced
scaling functions.

\subsubsection{SS-C transition, $|\Delta - \Delta_z| \ll \Delta$, $r>0$}
\label{ssc}

In this region we can neglect $\phi_z$ fluctuations and focus only on the
$\phi_x + i \phi_y$ which is undergoing Bose condensation. Further, it can
also be
shown that the second-order time derivative in ${\cal S}_{\phi}$ can be dropped.
Making these approximations, and defining
\begin{equation}
\Psi = \frac{\phi_x + i \phi_y}{\sqrt{\Delta_z}},
\end{equation}
we see that ${\cal S}_{\phi}$ reduces to
\begin{equation}
{\cal S}_{\Psi} = \int d^2 x \int_0^{1/T} d \tau
\left[ \Psi^{\ast} \frac{\partial \Psi}{\partial \tau}
+ \frac{c^2}{2 \Delta_z} |\nabla_x \Psi|^2 + (\Delta - \Delta_z) |\Psi|^2 +
\frac{u_0}{24 \Delta_z^2} |\Psi |^4 \right].
\label{hpsi}
\end{equation}
This action has been previously studied in some detail~\cite{popov,sss}: it has
a $z=2$ quantum critical point at $\Delta = \Delta_z$, and we will use the
existing
results later. Thus the SS-C transition is a line of $z=2$ critical points
terminating
in $z=1$ critical end-point M.

\subsubsection{N-C transition, $B \ll \rho_s (0)$, $r<0$}
\label{nc}

Both the $N$ and $C$
 phases are ordered, and it is sufficient to simply focus on static, thermal,
orientational fluctuations of the order parameter. We therefore quench the
magnitude
fluctuations of $\vec{\phi}$ and return to the fixed length vector $\vec{n}$.
The effective action for static $\vec{n}$ fluctuations
can be deduced from ${\cal S}_{\phi}$ to be
\begin{equation}
{\cal S}_n = \frac{1}{2 T} \int d^2 x \left[
\rho_s (0) ( \nabla_x \vec{n} )^2 + \gamma n_z^2 \right].
\label{hn}
\end{equation}
As noted earlier, $\rho_s (0)$ is the spin stiffness of the N\'{e}el state,
fully renormalized by quantum fluctuations. The anisotropy
$\gamma = 6 \Delta_z^2 (\Delta_z^2 - r)/u_0$ to lowest
order in $u_0$, and we expect $\gamma \sim \Delta_z^2$ more generally.
The action ${\cal S}_n$ has been studied in Ref~\onlinecite{np}, and we
will use their results in the following subsection.

\subsection{Non-zero temperature response functions}
\label{nonzerot}
A number of new phenomena occur at non-zero temperatures, and these are
addressed
in a relatively straightforward manner using the present continuum
effective field theory approach.
\begin{itemize}
\item
There is a broken $x-y$ symmetry in the C phase, and therefore a non-zero
temperature ($T_c$)
at which this order disappears in a Kosterlitz-Thouless transition.
An estimates $T_c$
was given earlier (sections \ref{KT} and Fig. \ref{fig33})
in the Hartree-Fock theory which
is  valid when the system is well away from one of the $T=0$ phase boundaries
of the C phase in Figs~\ref{mftfig}, and \ref{landau}.
We expect $T_c$ to vanish continuously as the system
in the C phase approaches the
$T=0$ boundaries to the N or the SS phase: there is nonzero temperature
phase transition above the N or the SS ground state. We discuss below the
behavior of $T_c$ near the C-N and C-SS $T=0$ phase boundaries. Near the
point M,
$T_c$ is determined completely and universally by the two energy scales which
measure the deviation of the ground state from M. So for $r>0$ we expect
\begin{equation}
T_c = \Delta_z  \Psi_{>} \left( \frac{\Delta}{\Delta_z } \right)
\label{tkt1}
\end{equation}
where $\Psi_{>}$ is a fully universal function; because the SS-C phase boundary
occurs precisely at $\Delta = \Delta_z $, we have
\begin{equation}
\Psi_{>} ( u \geq 1 ) = 0.
\end{equation}
Similarly for $r<0$ we have
\begin{equation}
T_c = \Delta_z  \Psi_{<} \left( \frac{\rho_s (0)}{\Delta_z } \right)
\label{tkt2}
\end{equation}
where $\Psi_{<}$ is also a universal function. Clearly the two functions
should agree
at $r=0$, and therefore we have $\Psi_{>} (0) =
\Psi_{<} (0)$; actually it is possible to say much more---for $\Delta_z
>0$ we expect that
$T_c$ is a smooth and analytic as a function of $r$ through $r=0$, and so using
the dependencies of $\Delta $ and $\rho_s (0)$ on $r$, it is possible to express
$\Psi_{>,<}$ as analytic continuations of each other.
We will give explicit expressions for $\Psi_{>,<}$ to leading order
in the expansion in $\epsilon=3-d$
below.
\item
The one-loop $T=0$ results for the SS phase (\ref{chiteq0})
predict infinitely sharp absorption peaks in $\chi_{\parallel}$ at $\omega
= \Delta$,
and in $\chi_{\perp}$ at $\omega = \Delta \pm \Delta_z $. As the SS phase
has a spin gap,
we expect these infinitely sharp peaks to survive at higher orders in the
perturbation
theory at $T=0$. For $T>0$ two qualitatively new features will arise.
First, thermal
damping will lead to a broadening of the peaks. Second, the peak positions will
themselves become temperature dependent. We will describe these processes below
in the vicinity of the point M, where both the broadening and the
$T$-dependent shifts
are quite significant. Deep inside the SS phase, well away from the M
point, these
$T$-dependencies are exponentially activated, and therefore much weaker.
\end{itemize}

We will restrict our results for the most part
to the paramagnetic phase, although results in the magnetically
ordered phases can be obtained by very similar methods. This means that we
are working
at $T>0$ above the SS phase, and at $T > T_c$ above the C phase, all within
the vicinity of the point M. The results are obtained using methods
discussed in some
detail in Ref~\onlinecite{ss2}: the only change is that the Zeeman splitting
$\Delta_z$ has
to be included in the propagators for the $\phi_{x,y}$ fields, and this
modifies the values
of the Matsubara frequency summations in the loop diagrams by replacing an
energy
$\varepsilon$ by $\varepsilon \pm \Delta_z $. The reader may also consult
Appendix D of 
Ref~\onlinecite{ds} where a simpler derivation of just the one loop results
of 
Ref~\onlinecite{ss2} is given.

The non-zero $T$ generalization of (\ref{chiteq0}) takes the form
\begin{eqnarray}
\chi_{\parallel}(\omega ) &=& \frac{1}{-\omega^2 + m_{\parallel}^2 - i
\Gamma_{\parallel}(\omega 
)} \nonumber \\
\chi_{\perp}(\omega ) &=& \frac{2}{-(\omega-\Delta_z )^2 + m_{\perp}^2 - i
\Gamma_{\perp}(\omega )} 
\nonumber
\end{eqnarray}
Here $m_{\parallel,\perp}$ and $\Gamma_{\parallel,\perp}$ depend implicitly
upon the
energy scales $T$, $\Delta_z $, and $\Delta$ ($\rho_s (0)$) for $r>0$
($r<0$) in a manner
we shall describe below to lowest order in $\epsilon$.
Clearly, the ``masses'' $m_{\parallel,\perp}$ represent the peak absorption
frequency,
while $\Gamma_{\parallel,\perp}$ are the absorptive pieces which lead to a
$T$-dependent broadening of the line.

First we describe the behavior of $m_{\perp,\parallel}$.

For $r>0$, the masses are universal functions $\Delta$, $T$ and $\Delta_z
$. They can be written
as
\begin{eqnarray}
m_{\parallel}^2 &=& R_{\parallel} - \epsilon \frac{2 \pi T}{11} \left(
3 \sqrt{R_{\parallel}} + 2 \sqrt{R_{\perp}} \right)
\nonumber \\
m_{\perp}^2 &=& R_{\perp} - \epsilon \frac{2 \pi T}{11} \left(
\sqrt{R_{\parallel}} + 4 \sqrt{R_{\perp}} \right)
\label{resm}
\end{eqnarray}
where
\begin{eqnarray}
R_{\parallel} &=& \Delta^2 \left[ 1 + \frac{5\epsilon}{11} \ln
\frac{T}{\Delta} \right] + 
\frac{\epsilon T^2}{11} \left[ 3 {G}\left(\frac{\Delta^2}{T^2}, 0\right) + 2
{G}\left(\frac{\Delta^2}{T^2}, \frac{\Delta_z }{T} \right) \right] \nonumber \\
R_{\perp} &=& \Delta^2 \left[ 1 + \frac{5\epsilon}{11} \ln \frac{T}{\Delta}
\right]
 + \frac{\epsilon T^2}{11} \left[  {G}\left(\frac{\Delta^2}{T^2}, 0\right) + 4
{G}\left(\frac{\Delta^2}{T^2}, \frac{\Delta_z }{T} \right) \right].
\label{resR}
\end{eqnarray}
The function ${G}(y,h)$ represents the value of the one-loop momentum integral;
it was computed in Refs~\onlinecite{ss2,ds} for the zero magnetic field
case $h=0$.
The generalization to non-zero $h$ is
\begin{equation}
{G}(y,h) = -2 \int_0^{\infty}  dq \left[ \ln \left( 2 q^2
\frac{\cosh ( \sqrt{q^2 + y}) - \cosh h}{q^2 + y - h^2} \right)
- q - \frac{y}{2 \sqrt{q^2 + 1/e}} \right]
\label{resG}
\end{equation}
This integral has to be evaluated numerically in general, but
we have the limiting value
${G}(0,0)= 2 \pi^2 / 3$.
Stability of the paramagnetic state requires that $m_{\perp} \geq
\Delta_z$; this requirement
leads to an expression for $T_c$, which is determined by solving $m_{\perp}
= \Delta_z $.
Analysis of this equation in powers of $\epsilon$ shows that $T_c \sim
1/\sqrt{\epsilon}$.
This implies that $\Delta/T , \Delta_z /T \sim \sqrt{\epsilon}$, and so
to leading order we can just use the value of $G(0,0)$ in (\ref{resR})
to obtain
\begin{equation}
T_c^2  =  \frac{33 ( \Delta_z ^2 - \Delta^2 )}{10 \pi^2 \epsilon}
\label{tc}
\end{equation}
for $\Delta_z  > \Delta$. For $\Delta_z < \Delta$ the system is in the SS phase,
and therefore $T_c =0$.
Notice that (\ref{tc}) agrees with the scaling form (\ref{tkt1}).
This result is expected to be the leading order result in powers of $\epsilon$,
except in the region $|\Delta_z  - \Delta| \ll \Delta$ where the $\epsilon$
expansion fails
and the reduced action ${\cal S}_{\Psi}$ of Section~\ref{ssc} has to be used.
Using results of Ref~\onlinecite{popov} for the latter action we have the exact
asymptotic form
\begin{equation}
T_c = \frac{(\Delta_z -\Delta) \ln [ \Delta_z /(\Delta_z -\Delta)]}{4 \ln
\ln [\Delta_z /(\Delta_z -\Delta)]}~~~~~~
\mbox{for $\ln[\Delta_z /(\Delta_z -\Delta)] \gg 1$}
\end{equation}

Closely related results can be obtained for $r<0$. In this case, the masses
are universal functions of $\rho_s (0)$, $\Delta_z $ and $T$.
However, considerable ambiguity arises in the $\epsilon$ expansion
for the result because $\rho_s (0)$ does not simply have the dimensions
of energy for all $d$. The appropriate scaling variable~\cite{ss2} is
$(\rho_s (0))^{1/(d-1)}$,
and it is necessary to keep the full $1/(d-1)$ power, rather than expand it
in powers of $\epsilon$ in order to make the engineering dimensions of the
results come
our correct. This then leads to ambiguities as to precisely which numerical
factors
should be raised to the power $1/(d-1)$ and which to $1/2 + \epsilon/4$. A
convenient
choice, which leads to the most compact expressions is to define
\begin{equation}
\widetilde{\rho_s} \equiv \left( \frac{2 \epsilon}{(n+8)} \frac{\rho_s}{S_{d+1}}
\right)^{1/(d-1)}
\end{equation}
where we have written the general expression for the $n$-component order
parameter: in the present case $n=3$. The factor $S_{d+1}$ is a phase-space
factor and is given by $S_d = 2/[ \Gamma(d/2) (4 \pi )^{d/2}]$ (this factor
was inadvertently omitted in Ref~\onlinecite{ss2}).
Notice that $\widetilde{\rho_s}$ has the dimensions of energy in $d=2$ (which is
of interest here). The value of $\widetilde{\rho_s}$, however must
be regarded as subject to large systematic corrections,
in view of the ambiguities noted above.
Using the methods and results of Ref~\onlinecite{ss2} for $r<0$,
and expressing them in terms of $\widetilde{\rho_s}$, we find that
the results (\ref{resm}) still hold, but (\ref{resR}) are replaced by
\begin{eqnarray}
R_{\parallel} =&& - \frac{\widetilde{\rho}_s^2 }{2} \left[ 1
-\frac{\epsilon}{22}
+ \frac{5\epsilon}{11} \ln \frac{T}{\widetilde{\rho}_s } \right]
+ \frac{\epsilon T^2}{11} \left[ 3 {G}\left(- \frac{\widetilde{\rho}_s^2
}{2 T^2},
0\right) + 2
{G}\left(- \frac{ \widetilde{\rho}_s^2 }{2 T^2}, \frac{\Delta_z }{T} \right)
\right] \nonumber \\
R_{\perp} =&& - \frac{\widetilde{\rho}_s^2 }{2}  \left[ 1
-\frac{\epsilon}{22}
+ \frac{5\epsilon}{11} \ln \frac{T}{\widetilde{\rho}_s } \right]
 + \frac{\epsilon T^2}{11} \left[  {G}\left(- \frac{ \widetilde{\rho}_s^2
}{2 T^2}, 0\right)
 + 4
{G}\left(- \frac{\widetilde{\rho}_s^2 }{2 T^2}, \frac{\Delta_z }{T} \right)
\right].
\end{eqnarray}
Notice that $G(y,h)$ is now needed for negative values of $y$.
Despite appearances, the expression (\ref{resG}) actually also holds for
$y<0$---one
simply uses the identity $\cosh(i x) = \cos(x)$ when the square root
becomes purely imaginary. Indeed, it is not difficult to show that the
expression
in (\ref{resG}) is actually analytic for all real $-\infty < y < \infty$
provided
$h > 0$.
We can use the same stability condition used for $r>0$ to now obtain
the leading order $\epsilon$-expansion result for $T_c$:
\begin{equation}
T_c^2  =  \frac{33 ( \Delta_z ^2 + \widetilde{\rho}_s^2 /2)}{10 \pi^2 \epsilon},
\label{tcrho}
\end{equation}
which is of the scaling form (\ref{tkt2}). The $\epsilon$ expansion fails
when $\Delta_z  \ll \rho_s (0)$ where the system approaches the C-N phase
boundary;
here, we use the effective action ${\cal S}_n$ of Section~\ref{nc},
and results for it in Ref~\onlinecite{np} to obtain
\begin{equation}
T_c = \frac{ 2 \pi \rho_s (0)}{\ln (\rho_s (0) / \Delta_z )}~~~~~~
\mbox{for $\ln ( \rho_s (0)/\Delta_z ) \gg 1$}.
\end{equation}

Finally, we obtain
the damping coefficients $\Gamma_{\perp,\parallel}$. This requires
evaluation of two-loop diagrams and the results are extremely lengthy.
We will be satisfied here by simply quoting the results valid
for $\Delta_z /T \ll 1$, $(\mbox{$\Delta$ or $\rho_s (0)$})/T \ll 1$ which were
obtained in Ref~\onlinecite{ss2}:
\begin{equation}
\Gamma_{\perp} ( \omega ) = \Gamma_{\parallel} ( \omega ) =
\frac{10 \pi \epsilon^2}{121} \left(
\frac{\omega^2}{8} + \pi^2 T^2 + 6 T^2 {\rm Li}_2 (e^{-\omega/2T}) \right)
\end{equation}
where ${\rm Li}_2 (x)$ is the dilogarithm function
\begin{equation}
{\rm Li}_2 ( x) = - \int_0^x \frac{dy}{y} \ln(1-y)
\end{equation}

\subsection{Connection to the Hartree-Fock theory}
The effective field theory for the double-layer
QH system at a filling factor of $\nu=2/m$
( with $m$ an odd integer)
that we develop above is entirely built on the
effective action $S_2$, defined by Eq. (\ref{action2}).
In particular, we make use of the fact that this effective
action for our problem is identical to the action of
the $2+1$ dimensional $O(3)$ quantum non-linear $\sigma$
model \cite{ss1,annals,ss3,ss2,popov,double,np,ds}
with the additional feature of an external magnetic field coupled
to the conserved global $O(3)$ charge. Once this precise mapping
of our effective action to that of the $2+1$ dimensional $O(3)$
quantum nonlinear $\sigma$ model becomes explicit,
the rest of the results derived in sections \ref{nls}A--C follow naturally.
The question now arises about the correspondence between our effective field
theory results in this section and the microscopic Hartree-Fock
results (for $\nu=2$) described in sections \ref{grou}--\ref{KT}.

It is to be noted that both the microscopic Hartree-Fock theory (sections
\ref{grou}--\ref{KT})
and the effective nonlinear $\sigma$ model field theory predict
the same number of zero temperature quantum phases, namely
the fully spin polarized ferromagnetic, the canted antiferromagnetic,
the N\'{e}el, and the symmetric spin singlet phase,
for the double-layer QH system at $\nu=2$. (The effective field
theory, in addition, enables us to predict that the double-layer
system at all fillings $\nu=2\nu_1$,
where $\nu_1=1/m$ with $m$ odd, has these four phases with the spin
singlet phase in the general case being
a non-trivial correlated SS phase rather than just the
pseudospin-symmetric spin-antisymmetric
SYM phase of the $\nu=2$ Hartree-Fock theory.)
It should also be noted that both the Hartree-Fock theory and
the effective field theory
predict the existence of a finite temperature Kosterlitz-Thouless
phase transition
in the canted antiferromagnetic phase with the planar antiferromagnetic ordering
disappearing above the Kosterlitz-Thouless transition temperature.
The underlying physics
in both the theories is that the system is essentially an $X$--$Y$
antiferromagnet in the layer (i.e. in the plane perpendicular to
the magnetic field direction) in the new canted phase.

On a more quantitative level it is easy to show that both theories predict
the same topology of the zero temperature quantum phase diagram.
This is demonstrated in Fig. \ref{fig15} where we have redrawn
the Hartree-Fock phase diagram (Fig. \ref{fig15}a) of Fig. \ref{fig21}
inverting abscissa (from $\Delta_{sas}/U_-$ to $U_-/\Delta_{sas}$)
and have somewhat reconfigured
the effective field theory phase diagram (Fig. \ref{fig15}b)
from Fig. \ref{mftfig} by keeping only the $QF_1$ phase
and by modifying the relative size of the various phases (which
are arbitrary within the effective field theory).
Using the definitions $t=(J/\rho_s^oM^2_o)^{1/2}$ from
Eq. (\ref{equ:tt1})
to define the abscissa, the effective field theory phase diagram in
$t-\Delta_z $
space (Fig. \ref{fig15}b) can be seen to be identical topologically to the
quantitatively calculated Hartree-Fock phase diagram (for $\nu=2$) in the
$\Delta_{sas}/U_--\Delta_z/\Delta_{sas}$
space (Fig. \ref{fig21}). Note that, in addition to the identical topology
involving four distinct quantum phases
as shown in Fig. \ref{fig15} of the two
phase diagrams with the effective coupling parameter $t$ of the field theory
(the abscissa in Fig. \ref{fig15}b) being proportional to
the parameter $\Delta_{sas}/U_-$ (the abscissa in Fig. \ref{fig15}a)
of the Hartree-Fock theory
(which is expected, because $t\sim \Delta_{sas}/U$ with $J$
being the interlayer superexchange coupling) and
the ordinate ($\sim \Delta_z $) being the same in both Figs. \ref{fig15}a
and \ref{fig15}b, the multicritical point M on the zero magnetic
field line shows up
in both phase diagrams. At the (zero temperature) quantum multicritical point M,
the canted, the spin singlet, and the N\'{e}el phase coexist. (The other
distinct multicritical point of the Hartree-Fock theory,
which is apparent on
the abscissa of Fig. \ref{fig21} where $\Delta_{sas}=\Delta_z$
and $d=0$, where the ferromagnetic, the canted and the symmetric phase
coexist is not accessible within the effective field theory because of its
long wavelength approximation, and can not be seen in
Fig. \ref{fig15}a as it is pushed to the point $U_-/\Delta_{sas}=0$,
$\Delta_z/\Delta_{\rm sas}=1$ where the two Hartree-Fock phase
boundaries of Fig. \ref{fig15}a come together.)
It is, therefore,
obvious that, except for very small values of $d$
(where the effective field theory which applies
only when $d>l_o$), the quantum phase diagrams predicted by the two theories
are topologically identical.

Finally, we can actually estimate the $\nu=2$ double-layer
Kosterlitz-Thouless transition temperature, $T_c$ of
section \ref{nls}C, in the effective field theory by using the
microscopic parameters obtained within the Hartree-Fock theory.
This calculation \cite{dsz1}, where one incorporate the
calculated Hartree-Fock parameters for $\Delta$ and $\epsilon=1$
in Eq. (\ref{tcrho}), leads~\cite{error} to an estimated effective
field theory $T_c \approx 3 K$
which is somewhat larger than the critical temperature
$T_c$ (Eq. (\ref{equ:t1})) estimated within the long wavelength mean
field Hartree-Fock treatment of section \ref{KT}.
In general, we believe the $\epsilon$--expansion leads substantial
overestimates of transition temperatures because it does
not properly account for the low-dimensional vortex effects responsible for
the transition.

\section{Comparison with experiments}
\label{exp}
In this section, we discuss some recent double-layer
$\nu=2$ inelastic light scattering experiments
whose findings are consistent with
our theoretical results.
A detailed quantitative comparison between our theory and the experiment
requires an accurate knowledge of
the temperature dependence of the related experimentally relevant response
functions as the system undergoes a finite temperature
phase transition at $T_c$.
Such a quantitative description is at present lacking,
and therefore we restrict ourselves mostly to a qualitative discussion.

In a recent inelastic light scattering experiment,\cite{ap1}
the long wavelength $\omega_0$ mode
of the intersubband SDW triplet
(see Fig. \ref{fig16} for schematic details of the various possible SDW
modes in the system),
which corresponds to transition
$|0\sigma\rangle\rightarrow|1\sigma\rangle$,
is measured for $\nu=2$
double-layer quantum Hall systems.
The double-layer samples used in the experiment
are by design in the canted antiferromagnetic phase
according to our zero temperature Hartree-Fock phase diagram,
{\it i.e.,} the ground state of the experimental system is the
canted antiferromagnetic quantum Hall state
(see Figs. \ref{fig3}, \ref{fig21}, and \ref{fig15} for the location
of the experimental sample in our theoretical diagram).
The experiment \cite{ap1} shows two important and striking features:
One is that there is a threshold temperature ($\sim0.5K$) below which the
$\omega_0$ mode becomes unobservable as it seems to lose
all spectral weight,
the other feature is that the excitation energy
$\omega_0$ approaches the Zeeman energy $\Delta_z$ when
the threshold temperature is approached from the above,
{\it i.e.} $\omega_0\approx\Delta_z$.
We argue below that these experimental observations
are completely consistent with our predicted Kosterlitz-Thouless
transition in the canted antiferromagnetic phase being the observed experimental
transition
at $T_c$.

First, we notice that the $\omega_0$ mode,
which involves a no-spin-flip transition with $\delta S_z=0$,
has a maximum spectral weight in the symmetric phase, where
there are as many spin-up (down) empty states as there are
spin-up (down) electrons.  The spectral weight of the $\omega_0$ mode
is identically
zero in the ferromagnetic phase, where all spin-up states
are occupied and all empty states are spin-down, and hence
the $\omega_0$ mode (which does not involve any spin flip) is forbidden.
The spectral weight of the $\omega_0$ mode should be nonzero but
small in the antiferromagnetic phase. This is because the
canted antiferromagnetic phase lies between the
symmetric phase and the
ferromagnetic phase in the phase diagram
and its spin-flip dynamics should thus be intermediate.
More over, the canted antiferromagnetic phase is not an eigenstate
for either spin or pseudospin, so the small spectral weight of the
$\omega_0$ mode is shared by many allowed transitions,
spreading the mode intensity over these transitions and thus making the
spectral weight of each transition even smaller.
It is thus plausible to regard the observed disappearance
of the $\omega_0$ mode
at the threshold temperature
as the transition to the
canted antiferromagnetic phase
at lower temperatures (where the spectral intensity for the $\omega_0$
mode becomes very small). Above the transition temperature
the system is essentially a disordered planar $X$-$Y$ magnet, and thus
behaves like a paramagnet whose SDW properties should be very similar to the
paramagnetic spin-singlet symmetric phase.

Next, we notice that,
in the symmetric phase,
the excitation
energies of the
intersubband SDW triplet
have the following simple relationship
\begin{equation}
\omega_\pm=\omega_0\pm\Delta_z.
\label{equ:3e}
\end{equation}
This expression can be derived explicitly, using either the
diagrammatic time-dependent Hartree-Fock approximation or the
single-mode approximation. It is a direct consequence
of the fact that Coulomb interaction is spin independent.
The above relationship bears a clear physical meaning:
$\omega_0\rightarrow\Delta_z$ means that
$\omega_-\rightarrow0$, {\it i.e.} mode softening (see Fig. \ref{fig16}).
Thus, the experimental observation that
$\omega_0$ approaches the Zeeman energy as the threshold temperature is
reached from above suggests that there is mode softening ($\omega_-=0$)
at the phase boundary,
as predicted by the computations of the $T$-dependent peak positions
in Section~\ref{nonzerot}.

Finally, we note that the
critical temperature (the threshold temperature)
in the experiment \cite{ap1}
is $T_c\approx0.52K$,
which is reasonably close to our calculated
Kosterlitz--Thouless critical temperature $T_c\approx1.8 K$
in the Hartree-Fock theory (Eq. (\ref{equ:t1}))
using the actual experimental sample parameters. \cite{dsz1}
This discrepancy between the experiment and the Hartree-Fock theory is
small when compared with the energy scale of
Coulomb interaction, which is about $70K$ in this
particular sample.
In addition,
quantum fluctuations neglected
in the Hartree-Fock theory should lower the
calculated critical temperature
and reduce this discrepancy.

{}From the above discussions, we conclude that our theoretical predictions
are consistent with the
recent light scattering
experimental results.
The most dramatic aspect of the experimental observations which give us
confidence in believing that the experiment is really seeing the canted
antiferromagnetic phase are (i) the
unambiguous observation of
a mode softening ({\it i.e.} $\omega_0\rightarrow\Delta_z$ implying
$\omega_-\rightarrow0$);
(ii) the observed temperature dependence
indicating a finite temperature phase transition;
(iii) the location of the experimental sample in
our calculated phase diagram and
(iv) the $\omega_0\rightarrow\Delta_z$ collapse being observed precisely
at $\nu=2$.

While the recent inelastic light scattering experiment \cite{ap1}
provides, in our opinion, rather compelling evidence in
favor of there being a finite temperature (Kosterlitz-Thouless)
transition in the $\nu=2$ double-layer system with the low
temperature phase being the canted antiferromagnetic phase (by virtue
of the vanishing of the $\omega_-$ mode at the phase boundary,
as discussed in section \ref{sdw} of this article), a complete verification
of our theory awaits further more conclusive and direct experimental
measurements, especially heat capacity measurements which should shown
(Fig. \ref{fig34}) power law temperature dependence in the canted
phase due to the existence of the Goldstone mode and exponential temperature
dependence in the two normal phases due to the existence of gaps
in the excitation spectra,
would be particularly well-suited in verifying our phase diagram.
The direct observation of a gapless Goldstone mode (Fig. \ref{fig7})
in the inelastic light scattering measurement in the (low temperature)
canted phase would also be rather definitive in establishing the
existence of the canted phase.
In this context we mention that the SDW softening indicating a phase
transition to the canted phase is a long wavelength instability,
and therefore optical spectroscopy \cite{gold1}
may also be useful in studying our proposed $\nu=2$ double-layer
phase diagram.
Both of these proposed direct experiments are fraught with
considerable (experimental) difficulties, however.
Electronic heat capacity measurements in quantum well structures
are notoriously difficult by virtue
of the extremely small magnitude of the (2D) electronic heat capacity
compared with the background
(lattice) contribution. As for the direct experimental observation
of the Goldstone mode, the experimental inelastic light scattering spectroscopy
is severely restricted by the selection rules inherent in the
resonant light scattering spectroscopy, and at this stage it is
unclear whether the problems associated with
the selection rules would allow to directly observe the Goldstone
mode.

One striking difference between the physics of $\nu=2$ double-layer
system and the corresponding $\nu=1$ situation is the existence
of a charge gap in the $\nu=2$ case for all values of $d$ and
$\Delta_{\rm sas}$: the system is always incompressible
(in all its quantum phases including the canted phase).
Thus the quantized Hall effect exists throughout our phase diagram unlike
in the corresponding $\nu=1$ situation. \cite{girvin1,girvin2,nu1,boe1}
The existence/nonexistence of the QH effect, which has been useful
in mapping out the $\nu=1$ double-layer phase diagram \cite{boe1}
would not work in our problem in a direct sense.
We do, however, speculate that the activation energy
({\it i.e.} the effective value of the incompressible charge gap)
for the $\nu=2\nu_1$ double-layer QH effect may very well
show observable structure at our calculated phase boundaries
even though all the phases
(ferromagnetic, canted, symmetric) would
exhibit $\nu=2\nu_1$ QH effect.
We suggest systematic experimental investigations of
$\nu=2\nu_1$ double-layer
( $\nu_1=1/m$ with $m=1,3,5,...$)
QH activation energies by tuning $\Delta_{\rm sas},\
\Delta_z$, and $d$ to look for signatures of our
proposed zero and finite temperature phase transitions.

In this context we point out that there is already some experimental
evidence \cite{pc1,pc2}
that the naive $\Delta_z=\Delta_{\rm sas}$ level crossing
in $\nu=2$ double-layer
QH systems does not exist (as our theory proposes and clearly
demonstrates in our calculated phase diagrams). The experimental observation
\cite{pc1,pc2} has been that
the naive $\nu=2$ level crossing phenomenon (at $\Delta_z=\Delta_{\rm sas}$)
between ferromagnetic and symmetric phases,
which would exist in the absence of our intervening canted phase,
if it happens at all in double-layer systems, must happen at
magnetic fields {\sl much lower} than that satisfying
$\Delta_z=\Delta_{\rm sas}$
condition.
This is, of course, exactly what our phase diagram
(see Fig. \ref{fig3})) predicts --- nothing interesting happens
for finite $d$ at $\Delta_z=\Delta_{\rm sas}$
or for that matter even for $\Delta_{\rm sas}=3 \Delta_z$
at $d=2l_o$ in Fig. \ref{fig3}a for example---
the system remains in the fully spin polarized ferromagnetic
phase and the naive expectation of a level crossing transition to the
symmetric phase simply does not occur.
In this sense, our phase diagram for
the $\nu=2$ double-layer system may have already been verified in
1990! \cite{boe1}
Further experiments along this line at $\nu=2\nu_1$ double-layer
systems would be useful.

\section{summary}
\label{sum}
In summary,
we have studied both zero and finite temperature
properties of the $\nu=2$
double-layer QH systems within the framework of
Hartree-Fock approximation.
We show that, in addition to the fully polarized state adiabatically
connected to the well separated layer state,
there are two other double-layer
quantum Hall phases: the first is a spin singlet,
and the second is characterized by a finite interlayer inplane
canted antiferromagnetic spin ordering.
The transition between the different quantum Hall phases
is continuous, and is signaled by the softening of
collective intersubband spin density excitations.
Because of the broken $U(1)$
symmetry in the canted antiferromagnetic phase,
the system has a finite temperature Kosterlitz-Thouless transition
($T_c\sim1K$).
Below the critical temperature,
the canted antiferromagnetic phase supports a linear Goldstone mode.
Above, the system is essentially a paramagnet similar
to the symmetric phase.
Our findings are consistent with recent
light scattering spectroscopic experimental results.
We present detailed results of our study,
including the stability energetics of various phases, the
antiferromagnetic order parameter
in the canted phase, the phase diagram,
the collective excitation dispersions,
the specific heat,
and the Kosterlitz-Thouless critical temperature,
and suggest various experiments which could, in principle,
probe the rich double-layer phase diagram predicted by our theory.

In addition to the microscopic $\nu=2$
Hartree-Fock theory, we have developed
a rather general long wavelength effective field theory for the $\nu=2\nu_1$,
where $\nu_1=1/m$ with $m$ an odd integer, double-layer
system. The essential inputs for
this effective field theory are the existence of charge gaps in the two
layers and an effective interlayer antiferromagnetic (superexchange)
interaction.  By mapping the effective action for this problem
to that of an $O(3)$ quantum nonlinear
$\sigma$ model, we have been able to show that the qualitative
phase diagram calculated within
the Hartree-Fock theory for $\nu=2$ is actually generically valid
(topologically) for any $\nu=2\nu_1$ (with $\nu_1=1,1/3,1/5,...$)
double-layer system with the symmetric phase of the Hartree-Fock calculation
being replaced by a highly non-trivial correlated spin singlet
phase ( of which the $\nu=2$ symmetric phase is a rather trivial example).
Thus, there could be rather non-trivial canted (and perhaps even N\'{e}el,
if one can apply sufficient external pressure to produce
vanishing gyromagnetic ratio)
antiferromagnets at, for example, $\nu=2/3$ in a double-layer
system, where each single fully spin polarized
Laughlin state spontaneously develops in-plane antiferromagnetic spin
ordering. Observation of the canted or the spin-singlet phase in a
$\nu=2/3$ double-layer QH system would significantly enrich
the many-body strong correlation physics associated with QH systems.

We conclude by pointing out that, although we have confined ourselves
in this article to the $\nu=2/m$ case, with $m$ an odd integer,
it is obvious that the physics we are considering here applies,
in principle, to all double-layer QH systems with $\nu=2\nu_1$
where a single layer at filling $\nu_1$ forms
a fully {\sl spin polarized}
incompressible QH state with a charge gap. Thus, the same physics
as at $\nu=2$ should
apply, in principle, at $\nu=6$ (but {\sl not} at $\nu=4,8,...$
where the charge gap is the cyclotron gap not $\Delta_z, \Delta_{\rm sas}$.)
in double-layer system.
In principle, however, our approximations which neglect
all (orbital) Landau level
couplings become progressively
worse at higher Landau levels. In this respect, it is very gratifying
that the experimental light scattering
measurements \cite{ap1} find qualitatively similar
(but quantitatively much suppressed)
behavior at $\nu=6$ as at $\nu=2$, but the $\nu=4$ situation is qualitatively
different.

\acknowledgments
The authors
thank Dr. A. Pinczuk and Dr. V. Pellegrini
for helpful discussions on the experimental data.
The work at University of Maryland is supported by
the U.S.-ONR and the U.S.-ARO.
The work at Yale University is supported by NSF Grant No
DMR 96-23181.

\appendix

\section{Two spin problem}
\label{twospin}
Here we will assess the validity of the mapping from the action ${\cal S}_1$
in (\ref{action1}) to ${\cal S}_2$ in (\ref{action2}) by examining a simple
toy model of two spins. We consider the Hamiltonian
\begin{equation}
{\cal H} = J \vec{S}_1 \cdot \vec{S}_2 - \Delta_z \hat{z} \cdot ( \vec{S}_1
+ \vec{S}_2 )
\end{equation}
where $\vec{S}_{1,2}$ are two quantum spins of spin $S$. The energy spectrum of
${\cal B}$ is clearly
\begin{equation}
E_{\ell} = \frac{J}{2} \ell ( \ell + 1) - \Delta_z  m + E_0~~~~~~~~,
~~~~~~~~\ell =0,1,\ldots 2 S~~;~~
m = -\ell, -\ell + 1, \ldots \ell-1, \ell
\label{app0}
\end{equation}
where $E_0$ is an overall constant we shall  not be interested in.

Let us attempt to obtain this result using the coherent state path integral.
First, we transcribe ${\cal H}$ into the effective action
\begin{equation}
{\cal S} = \int d \tau \left[ i S \vec{A}(\vec{n_1}) \cdot
\frac{\partial \vec{n_1 }}{\partial \tau} + i S \vec{A}(\vec{n_2}) \cdot
\frac{\partial \vec{n_2 }}{\partial \tau} + J S^2 \vec{n}_1 \cdot \vec{n}_2
\right]
\label{app1}
\end{equation}
where $\vec{n}_{1,2}^2 = 1$.
Notice that this is the analog of the action ${\cal S}_1$ in (\ref{action1})
with only the spatial gradient spin stiffness terms now being absent.
Now insert the parameterization (\ref{nl}) into (\ref{app1}), and expand
to quadratic order in $\vec{L}$. The neglect of terms higher order in
$\vec{L}$ is the only approximation being made here.
This gives us the analog of (\ref{action1a})
\begin{equation}
{\cal S} \approx \int d \tau
\left[ 2 i S \vec{L} \cdot\left( \vec{n} \times
\frac{\partial \vec{n}}{\partial \tau} + i \Delta_z \hat{z} \right) + 2 J
\vec{L}^2
\right]
\label{app2}
\end{equation}
Now we integrate out $\vec{L}$ as described above (\ref{action2}) to obtain
\begin{equation}
{\cal S} \approx \int d \tau
\frac{1}{2 J} \left( \frac{\partial
\vec{n}}{\partial \tau}
- i \Delta_z \hat{z} \times \vec{n} \right)^2
\end{equation}
where recall that the functional integral is over the unit vector field
$\vec{n} ( \tau)$ satisfying $\vec{n}^2 = 1$ for all $\tau$.
This last form of ${\cal S}$ is the effective action for a quantum
rotor in a field $\Delta_z \hat{z}$. This action is equivalent to the
Hamiltonian
\begin{equation}
{\cal H}_R = \frac{J}{2} \hat{\vec{L}}^2 - \Delta_z \hat{z} \cdot \hat{\vec{L}}
\end{equation}
where $\hat{\vec{L}}$ is the rotor angular momentum operator. The eigenvalues of
${\cal H}_R$ are easily seen to be identical to those of ${\cal H}$ in
(\ref{app0}) with one simple difference. The allowed values of
$\ell$ now extend over {\em all} non-negative integers. Thus the only effect
of dropping terms higher order in $\vec{L}$ in the functional analysis
is that the upper bound $\ell \leq 2S$ has disappeared. This only
introduces additional
states at relatively high energies and is therefore not expected to be of
importance
in our study of the low energy properties of ${\cal S}_2$.

\begin{figure}
\caption{
The energy per magnetic flux in
the symmetric (SYM) state,
the spin polarized ferromagnetic (FM) state, and
the canted antiferromagnetic (C) state for
a $\nu=2$ double-layer system with
$\Delta_{\rm sas}=0.07e^2/\epsilon l_o$,
$\Delta_z=0.01e^2/\epsilon l_o$,
and the well-thickness $d_w=0.8l_o$.
}
\label{fig1}
\end{figure}

\begin{figure}
\caption{
Schematic display of electron spin orientations in the canted
antiferromagnetic quantum Hall phase.
}
\label{fig41}
\end{figure}

\begin{figure}
\caption{
The canted
antiferromagnetic order parameter
versus layer separation for the indicated tunneling
and Zeeman energies.
The well-thickness $d=0.8l_o$.
}
\label{fig2}
\end{figure}

\begin{figure}
\caption{
The zero temperature phase diagrams at $\nu=2$ within
the Hartree-Fock approximation at two different values of the
Zeeman energy: (a) $\Delta_z=0.01e^2/\epsilon l_o$
and (b) $\Delta_z=0.01e^2/\epsilon l_o$.
The quantum well thickness is $d_w=0.8l_o$
for both the figures.
Three phases are present: the symmetric phase (SYM),
the spin polarized
ferromagnetic phase (FM), and the
canted antiferromagnetic phase (C).
The `+' in (a) denotes the experimental sample parameters
of Ref.\ {\protect\onlinecite{ap1}}.
The vertical dotted line
in each figure indicated the $\Delta_z=\Delta_{\rm sas}$ condition,
which is the naive phase boundary between the FM
($\Delta_z>\Delta_{\rm sas}$) and the SYM
($\Delta_z<\Delta_{\rm sas}$) phase
with an expected level crossing at $\Delta_z=\Delta_{\rm sas}$.
}
\label{fig3}
\end{figure}

\begin{figure}
\caption{
Feynman diagram for the intersubband spin density response function
in the time-dependent Hartree-Fock approximation,
where solid lines are the self-consistent Hartree-Fock
electron Greens functions and wiggled lines are Coulomb
interaction potentials.
}
\label{fig4}

\end{figure}
\begin{figure}
\caption{
The inter-subband SDW dispersion $\omega_{\pm}(q)$
in the spin polarized ferromagnetic (FM) phase at $\nu=2$
with tunneling energy $\Delta_{\rm sas}=0.02 e^2/\epsilon l_o$,
Zeeman energy $\Delta_z=0.01e^2/\epsilon l_o$,
layer separation $d=1.15l_o$,
and the well-thickness $d_w=0.8l_o$.
}
\label{fig5}

\end{figure}
\begin{figure}
\caption{
The inter-subband SDW dispersion $\omega_{\pm}(q)$
in the symmetric (SYM) phase at $\nu=2$ with
layer separation $d=0.85l_o$,
Zeeman energy
$\Delta_{z}=0.08e^2/\epsilon l_o$,  tunneling energy
$\Delta_{\rm sas}=0.35e^2/\epsilon l_o$,
and the well-thickness $d_w=0.8l_o$.
}
\label{fig6}
\end{figure}

\begin{figure}
\caption{
The low energy intersubband SDW mode $\omega_-(q=0)$
and the canted antiferromagnetic order parameter (COP)
versus tunneling energy with
layer separation $d=1.0l_o$,
Zeeman energy
$\Delta_z=0.08e^2/\epsilon l_o$,
and the well-thickness $d_w=0.8l_o$.
}
\label{fig8}
\end{figure}

\begin{figure}
\caption{
The inter-subband SDW dispersion $\omega_{\pm}(q)$
in the canted antiferromagnetic (C) phase at $\nu=2$ with
layer separation $d=1.15l_o$,
tunneling energy
$\Delta_{\rm sas}=0.14e^2/\epsilon l_o$, Zeeman energy
$\Delta_z=0.01e^2/\epsilon l_o$,
and the well-thickness $d_w=0.8l_o$.
}
\label{fig7}
\end{figure}

\begin{figure}
\caption{
Zero temperature phase diagram of a $\nu=2$ double-layer quantum
Hall system within the Hartree-Fock approximation.
The phase diagram is expressed in terms of
scaled dimensionless variables.
The `+' mark represents the experimental sample of Ref.
{\protect\onlinecite{ap1}}.
The N\'{e}el phase (N) at
$\Delta_z=0$ and $\Delta_{\rm sas}<2U_-$ is represented by the thick line.
The M-point represents the quantum critical point at $\Delta_z=0$.
}
\label{fig21}
\end{figure}

\begin{figure}
\caption{
The calculated Kosterlitz--Thouless critical temperature $T_c$
versus tunneling energy $\Delta_{\rm sas}$
at different
interlayer separations: dotted line $d=1.4\ l_o$,
solid line $d=1.2\ l_o$, and dashed line $d=1.0\ l_o$.
Zeeman energy $\Delta_z=0.04\ e^2/\epsilon l_o$.
The layer-thickness $d_w=0.8\ l_o$.
}
\label{fig33}
\end{figure}
\begin{figure}
\caption{
The heat capacity per magnetic flux
of a $\nu=2$ double-layer quantum Hall system
as functions of temperature
in the symmetry phase (SYM), in the spin-polarized ferromagnetic phase (FM),
and in the canted antiferromagnetic phase (C).
The inset shows $C/{\cal T}^2$, where
${\cal T}=k_BT/(e^2/\epsilon l_o)$, versus $T$
in the C phase.
}
\label{fig34}

\end{figure}

\begin{figure}
\caption{Mean field phase diagram of the quantum rotor Hamiltonian ${\cal H}$
in (\protect\ref{hrotor}). The phases are described in
Section~\protect\ref{mft}.
Only the ${\rm QF}_1$ phase is expected to appear for the two-layer model under
consideration here, and is referred to elsewhere as the FM: the ${\rm
QF}_2$ phase is an 
artifact of the approximations made in
deriving the rotor model.
The SS phase was also called the SYM phase in the Hartree Fock computations.
}
\label{mftfig}
\end{figure}

\begin{figure}
\caption{Mean field phase diagram of the soft-spin action ${\cal S}_{\phi}$
in (\protect\ref{hphi}). The SS phase was also called the SYM phase in the
Hartree Fock computations. Notice that it captures the vicinity of the point
M in the rotor mean field phase diagram in Fig~\protect\ref{mftfig}.
The multicritical point M is described by a relativistic continuum field theory
with dynamic exponent $z=1$. The SS-C boundary is a line of second order
transitions
with dynamic exponent $z=2$ and is described by action ${\cal S}_{\Psi}$
in (\protect\ref{hpsi}).
The position of this boundary is given exactly by $\Delta = \Delta_z$,
where $\Delta \sim
r^{\nu}$ is the $\Delta_z=0$ spin gap of the SS phase ($\nu$ is the
correlation length exponent
of M). The action ${\cal S}_{\Psi}$ holds for $|\Delta -\Delta_z |
\ll \Delta$.
The N state has $T=0$ spin stiffness $\rho_s (0) \sim (-r)^{\nu}$,
and for $\Delta_z \ll \rho_s (0)$, the action ${\cal S}_{n}$ in
(\protect\ref{hn})
describes low $T$ fluctuations.}
\label{landau}
\end{figure}

\begin{figure}
\caption{
(a) The Zero temperature phase diagram of a double-layer
quantum Hall system at $\nu=2$ within the Hartree-Fock approximation.
This is the same diagram as Fig.~\protect\ref{fig21}. It is redrawn here
with the abscissa inverted.
The `+' mark represents the experimental sample of Ref.
{\protect\onlinecite{ap1}}.
The N\'{e}el phase (N) at
$\Delta_z=0$ and $\Delta_{\rm sas}<2U_-$ is represented by the thick line.
(b) Zero temperature phase diagram of a double-layer quantum Hall system
at $\nu=2\nu_1$ derived from the effective Lagrangean ${\cal S}_2$
(Eq. ({\protect\ref{action2}})).
The inset shows the topologically identical
Hartree-Fock phase diagram of Fig.~\protect\ref{fig21}.
The FPF, C, and SS phases in the main figure correspond to the
FM, AF, and SYM phases in the inset, respectively.
}
\label{fig15}
\end{figure}
\begin{figure}
\caption{
The intersubband spin excitation transitions
in a double-layer quantum Hall system at $\nu=2$
in the (a) symmetric phase, (b) ferromagnetic phase, and (c)
the canted antiferromagnetic phase.
The spin conserved transition
($\omega_0$ mode) has large spectral weight in the symmetric phase
and is prohibited in the ferromagnetic phase.
}
\label{fig16}
\end{figure}


\begin{references}
\bibitem{rk1}
The finite well-thickness corrections, as well as other detailed
corrections to interaction potentials, have quantitative effects only.
They do not change any results qualitatively.
\bibitem{rk3}
For notational clarity, we use $\mu$ for the subband index,
$L$ ($R$) for the layer index, $\sigma$ for the electron spin,
$\alpha$ for the intra-Landau level index,
and $i\lambda$ ($i=1,2$ and $\lambda=\pm$) for
the four Hartree-Fock eigenstates.
\bibitem{error} The estimate of $T_c$ towards the conclusion of
Ref~\protect\onlinecite{dsz1} is in error because of the omission
noted in~\protect\onlinecite{ss2}.
\end{references}
\end{document}